\documentclass[aps,amsfonts,amsmath,prd,preprint,nofootinbib]{revtex4}
\pdfoutput=1
\usepackage{graphicx}

\newcommand{\beq}{\begin{equation}}
\newcommand{\eeq}{\end{equation}}

\def\lap{\lower.5ex\hbox{$\; \buildrel < \over \sim \;$}}
\def\gap{\lower.5ex\hbox{$\; \buildrel > \over \sim \;$}}

\def\be{\begin{equation}}
\def\ee{\end{equation}}
\def\ba{\begin{eqnarray}}
\def\ea{\end{eqnarray}}

\begin{document}

\title{Black holes and the multiverse}

\author{Jaume Garriga$^{a,b}$, Alexander Vilenkin$^b$ and Jun Zhang$^b$}

\address{$^a$ Departament de Fisica Fonamental i Institut de Ciencies del Cosmos,\\
Universitat de Barcelona, 
Marti i Franques, 1, 08028, Barcelona, Spain\\
$^b$Institute of Cosmology, Department of Physics and Astronomy,\\ 
Tufts University, Medford, MA 02155, USA}

\begin{abstract}

Vacuum bubbles may nucleate and expand during the inflationary epoch in the early universe. After inflation ends, the bubbles quickly dissipate their kinetic energy; they come to rest with respect to the Hubble flow and eventually form black holes. The fate of the bubble itself depends on the resulting black hole mass. If the mass is smaller than a certain critical value, the bubble collapses to a singularity.  Otherwise, the bubble interior inflates, forming a baby universe, which is connected to the exterior FRW region by a wormhole.  A similar black hole formation mechanism operates for spherical domain walls nucleating during inflation. As an illustrative example, we studied the black hole mass spectrum in the domain wall scenario, assuming that domain walls interact with matter only gravitationally.  Our results indicate that, depending on the model parameters, black holes produced in this scenario can have significant astrophysical effects and can even serve as dark matter or as seeds for supermassive black holes.  The mechanism of black hole formation described in this paper is very generic and has important implications for the global structure of the universe.  Baby universes inside super-critical black holes inflate eternally and nucleate bubbles of all vacua allowed by the underlying particle physics.  The resulting multiverse has a very non-trivial spacetime structure, with a multitude of eternally inflating regions connected by wormholes. If a black hole population with the predicted mass spectrum is discovered,
it could be regarded as evidence for inflation and for the existence of a
multiverse.

\end{abstract}

\maketitle

\section{Introduction}

A remarkable aspect of inflationary cosmology is that it attributes the origin of galaxies and large-scale structure to small quantum fluctuations in the early universe \cite{Mukhanov}.  The fluctuations remain small through most of the cosmic history and become nonlinear only in relatively recent times.  Here, we will explore the possibility that non-perturbative quantum effects during inflation could also lead to the formation of structure on astrophysical scales.  Specifically, we will show that spontaneous nucleation of vacuum bubbles and spherical domain walls during the inflationary epoch can result in the formation of black holes with a wide spectrum of masses.  

The physical mechanism responsible for these phenomena is easy to understand.  The inflationary expansion of the universe is driven by a false vacuum of energy density $\rho_i \approx {\rm const}$.  Bubble nucleation in this vacuum can occur \cite{CdL} if the underlying particle physics model includes vacuum states of a lower energy density, $\rho_b < \rho_i$.\footnote{Higher-energy bubbles can also be formed, but their nucleation rate is typically strongly suppressed \cite{LeeWeinberg}.}  We will be interested in the case when $\rho_b > 0$.  Once a bubble is formed, it immediately starts to expand.  The difference in vacuum tension on the two sides of the bubble wall results in a force $F = \rho_i - \rho_b$ per unit area of the wall, so the bubble expands with acceleration.  This continues until the end of inflation (or until the energy density of the inflating vacuum drops below $\rho_b$ during the slow roll).  
At later times, the bubble continues to expand but it is slowed down by momentum transfer to the surrounding matter, while it is also being pulled inwards by the negative pressure of vacuum in its interior. Eventually, this leads to gravitational collapse and the formation of a black hole. 

The nature of the collapse and the fate of the bubble interior depends on the bubble size.  The positive-energy vacuum inside the bubble can support inflation at the rate $H_b =(8\pi G\rho_b/3)^{1/2}$, where $G$ is Newton's constant.  If the bubble expands to a radius $R\gtrsim H_b^{-1}$, its interior begins to inflate\footnote{This low energy internal inflation takes place inside the bubble, even though inflation in the exterior region has already ended.}. We will show that the black hole that is eventually formed contains a ballooning inflating region in its interior, which is connected to the exterior region by a wormhole.   
On the other hand, if the maximum expansion radius is $R\ll H_b^{-1}$, then internal inflation does not occur, and the bubble interior shrinks and collapses to a singularity.\footnote{For simplicity, in this discussion we disregard the gravitational effect of the bubble wall.  We shall see later that if the wall tension is sufficiently large, a wormhole can develop even when $R\ll H_b^{-1}$.}

Bubbles formed at earlier times during inflation expand to a larger size, so at the end of inflation we expect to have a wide spectrum of bubble sizes.  They will form black holes with a wide spectrum of masses.  We will show that black holes with masses above a certain critical mass have inflating universes inside.

The situation with domain walls is similar to that with vacuum bubbles.  It has been shown in Refs. \cite{Basu} that spherical domain walls can spontaneously nucleate in the inflating false vacuum.  The walls are then stretched by the expansion of the universe and form black holes when they come within the horizon.  Furthermore, the gravitational field of domain walls is known to be repulsive \cite{AV83,Sikivie}.  This causes the walls to inflate at the rate $H_\sigma = 2\pi G\sigma$, where $\sigma$ is the wall tension.  We will show that domain walls having size  $R\gtrsim H_\sigma^{-1}$ develop a wormhole structure, while smaller walls collapse to a singularity soon after they enter the cosmological horizon.  Once again, there is a critical mass above which black holes contain inflating domain walls connected to the exterior space by a wormhole.

We now briefly comment on the earlier work on this subject.  Inflating universes contained inside of black holes have been discussed by a number of authors \cite{Kodama,Berezin,Blau}.  Refs.~\cite{Berezin,Blau} focused on black holes in asymptotically flat or de Sitter spacetime, while Ref.~\cite{Kodama} considered a different mechanism of cosmological wormhole formation.  The possibility of wormhole formation in cosmological spacetimes has also been discussed in Ref.~\cite{Carr21}, but without suggesting a cosmological scenario where it can be realized.  Cosmological black hole formation by vacuum bubbles was qualitatively discussed in Ref.~\cite{watcher}, but no attempt was made to determine the resulting black hole masses.  Black holes formed by collapsing domain walls have been discussed in Refs.~\cite{GV,Yoo,Khlopov}, but the possibility of wormhole formation has been overlooked in these papers, so their estimate of black hole masses applies only to the subcritical case (when no wormhole is formed).

In the present paper, we shall investigate cosmological black hole formation by domain walls and vacuum bubbles that nucleated during the inflationary epoch.  We shall study the spacetime structure of such black holes and estimate their masses.  We shall also find the black hole mass distribution in the present universe and derive observational constraints on the particle physics model parameters.  We shall see that for some parameter values the black holes produced in this way can serve  as dark matter or as seeds for supermassive black holes observed in galactic centers.

The paper is organized as follows.  Sections \ref{gcb} and \ref{gcdw} are devoted, respectively, to the gravitational collapse of vacuum bubbles and domain walls. Section \ref{mdbh} deals with the mass distribution of black holes produced by the collapse of domain walls, and Section \ref{ob} with the observational bounds on such distribution. Our conclusions are summarized in Section \ref{conclusions}. A numerical study of test domain walls is deferred to Appendix A, and some technical aspects of the spacetime structure describing the gravitational collapse of large domain walls are discussed in Appendix B.

\section{Gravitational collapse of bubbles} \label{gcb}

In this Section we describe the gravitational collapse of bubbles after inflation, when they are embedded in the matter distribution of a FRW universe.
Three different time scales will be relevant for the dynamics. These are the cosmological scale 
\begin{equation}
t \sim H^{-1} = (3/8\pi G\rho_m)^{1/2},
\end{equation}
where $\rho_m$ is the matter density,
the scale associated with the vacuum energy inside the bubble,\begin{equation}
t_b \equiv H_b^{-1} = (3/8\pi G\rho_b)^{1/2},
\end{equation}
and the acceleration time-scale due to the repulsive gravitational field of the domain wall
\begin{equation}
t_\sigma \equiv H_\sigma^{-1}=  (2\pi G\sigma)^{-1}.
\end{equation}
In what follows we assume that the the inflaton transfers its energy to matter almost instantaneously, and for definiteness we shall also assume a separation of scales,
\begin{equation}
t_i \ll t_b,t_\sigma, \label{separation}
\end{equation}
where $t_i\sim H_i^{-1}$ is the Hubble radius at the time when inflation ends. The relation (\ref{separation}) guarantees that the repulsive gravitational force due to the vacuum energy and wall tension 
are subdominant effects at the end of inflation, and can only become important much later. 

The interaction of bubbles with matter is highly model dependent.
For the purposes of illustration, 
here we shall assume that matter is created only outside the bubble, and has reflecting boundary conditions at the bubble wall. For the case of domain walls, which will be discussed in the following Section, we shall consider the case where matter is on both sides of the wall. As we shall see, this leads to a rather different dynamics.

\label{bubbles}

The dynamics of spherically symmetric vacuum bubbles has been studied in the thin wall limit by Berezin, Kuzmin and Tkachev \cite{Berezin}. 
\footnote{A more detailed and pedagogical treatment in the asymptotically flat case was later given by Blau, Guendelman and Guth \cite{Blau}.}
The metric inside the bubble is the de Sitter (dS) metric with radius $H_b^{-1}$. The metric outside the bubble is the Schwarzschild-de Sitter metric\footnote{Here we are ignoring slow roll corrections to the vacuum energy during inflation.} (SdS), with a mass parameter which can be expressed as
\begin{equation} 
M={4\over 3} \pi (\rho_b-\rho_i) R^3+ 4\pi \sigma R^2 [\dot R^2 +1-H_b^2 R^2]^{1/2} - 8\pi^2 G \sigma^2 R^3. \label{mass}
\end{equation}
Here, $\rho_i$ is the vacuum energy outside the bubble, $R$ is the radius of the bubble wall, and $\dot R\equiv dR/d\tau$, where $\tau$ is the proper time on the bubble wall worldsheet, at fixed angular position. Eq. (\ref{mass}) can  be interpreted as an energy conservation equation.
\footnote{In principle, both signs for the square root are allowed for the second term at the right hand side. However, for $\rho_b<\rho_i$ (which holds  during inflation, when the 
energy density in the parent vacuum is higher than in the new vacuum) only the
plus sign leads to a positive mass $M$, so the negative sign must be discarded.}

\subsection{Initial conditions} \label{subinitial}

A bubble nucleating in de Sitter space has zero mass, $M_{dS}=0$, so the right hand side of (\ref{mass}) vanishes during inflation. 
Assuming that $R$, $\dot R$ and 
the bubble wall tension are continuous in the transition from inflation to the matter dominated regime, we have
\begin{equation}
{4\over 3} \pi \rho_m(t_i) R_i^3 \approx {4\over 3} \pi \rho_b R_i^3 + 4\pi \sigma R_i^2 [\dot R_i^2 +1-H_b^2 R_i^2]^{1/2} - 8\pi^2 G \sigma^2 R_i^3. \label{massi}
\end{equation}
Here, we have assumed that the transfer of energy from the inflaton to matter takes place almost instantaneously at the time $t_i$, 
so that 
$\rho_m(t_i)\approx \rho_i$, and that the vacuum energy $\rho_0$ outside the bubble is completely negligible after inflation,
\begin{equation}
\rho_0 \ll \rho_m.
\end{equation}
Note that the left hand side of Eq. (\ref{massi}),
\begin{equation}
M_i \equiv {4\over 3}\pi \rho_m(t_i)R_i^3 \label{excluded}
\end{equation} 
is the ``excluded" mass of matter in a cavity of radius $R_i$ at time $t_i$, which has been replaced by a bubble of a much lower vacuum energy density $\rho_b$. The difference 
in energy density goes into the kinetic energy of the bubble wall and its self-gravity corrections, corresponding to the second and third terms in Eq. (\ref{massi}).

Let us now calculate the Lorentz factor of the bubble walls with respect to the Hubble flow at the time $t_i$. Assuming the separation of scales (\ref{separation}), the right hand side of Eq. (\ref{massi}) is dominated by 
the kinetic term and we have
\begin{equation}
\dot R_i \approx {1\over 4} H_i^2 R_i t_\sigma.\label{dotri}
\end{equation}
We wish to consider the motion of the bubble relative to matter. The metric outside the bubble is given by
\begin{equation}
ds^2 = -dt^2 + a^2(t) (dr^2 + r^2 d\Omega^2).
\end{equation}
Here, $a(t) \propto t^\beta$ is the scale factor, where $\beta=1/2$ for radiation, or
$\beta=2/3$ for pressureless matter, as may be the case if the energy is in the form of a scalar field oscillating around the minimum of a quadratic potential. The proper time 
on the worldsheet is given by 
\begin{equation}
d\tau = \sqrt{1-a^2r'^2} dt,
\end{equation}
where a prime indicates derivative with respect to $t$, and we have
\begin{equation}
{dR \over d\tau} = {HR + a r' \over \sqrt{1-a^2r'^2}},\label{dotri2}
\end{equation}
with $H=a'/a=\beta/t$. The typical size of bubbles produced during inflation is at least 
of order $t_i$, although it can also be much larger. Hence, at the end of inflation we have $H_iR_i \gtrsim 1$. 
On the other hand, the second term in the numerator
of (\ref{dotri2}) is bounded by unity, $ar'<1$. Combining (\ref{dotri}) and (\ref{dotri2}) we have
\begin{equation}
\gamma_i \equiv {1 \over \sqrt{1-a_i^2{r'}_i^2}} \sim {t_\sigma\over t_i}, \label{initial}
\end{equation}
and using (\ref{separation}), we find that the motion of the wall relative to matter is highly relativistic, $\gamma_i \gg 1$.
 
 \subsection{Dissipation of kinetic energy}\label{dissipation}
 
The kinetic energy of the bubble walls will be dissipated by momentum transfer to the surrounding matter. Let us now estimate the time-scale for this to happen. The force acting on the wall due to momentum transfer per unit area in the radial direction can be estimated as 
\begin{equation}
T^{\hat 0\hat r}\sim \rho_m U^{\hat 0}U^{\hat r} \sim - \rho_m \gamma^2, \label{transfer} 
\end{equation}
where $U^{\hat\mu}$ is the four-velocity of matter in an orthonormal frame in which the wall is at rest, hatted indices indicate tensor components in that frame, and we have used that the motion of matter is highly relativistic in the radial direction\footnote{ Here we use a simplified description where the momentum transfer is modeled by superimposing an incoming fluid with four velocity $U^{\hat\mu}$ that hits the domain wall, and a reflected fluid which moves in the opposite direction. This description should be valid in the limit of a very weak fluid self-interaction. In a more realistic setup, shock waves will form 
\cite{ML,BM}, but we expect a similar behaviour for the momentum transfer.}.
The corresponding proper acceleration of the wall is given by
\begin{equation}
\alpha \sim -{\rho_m \gamma^2 \over \sigma}. \label{alpha1}
\end{equation}
In the rest frame of ambient matter, we have 
\begin{equation}
{d\gamma \over dt} = \gamma^3 v v'= v \alpha \approx \alpha, \label{alpha2}
\end{equation}
where $v=a r'$ is the radial outward velocity of the wall. The second equality just uses the kinematical relation between acceleration and proper acceleration for the case when velocity and acceleration are parallel to each other. Combining (\ref{alpha1}) and (\ref{alpha2}) we have
\begin{equation}
{d\gamma \over \gamma^2} \sim - t_\sigma {dt\over t^2}. \label{de}
\end{equation}
Solving (\ref{de}) with the initial condition (\ref{initial}) at $t=t_i$, it is straightforward to show that 
the wall loses most of its energy on a time scale 
\begin{equation}
\Delta t \sim {t_i^3 \over t_\sigma^2} \ll t_i, \label{timescale}
\end{equation}
and that the wall slows down to $\gamma\sim 1$ in a time of order
$\Delta t' \sim {t_i^2 /t_\sigma} \ll t_i$, which is still much less than the Hubble time at the end of inflation.

Since the bubble motion will push matter into the unperturbed FRW region, this results in a highly overdense shell surounding the bubble, exploding at a highly relativistic speed.  The total energy of the shell is ${\cal E}_{shell}\approx M_i$, with most of this energy contained in a thin layer of mass $M_{shell}\sim (t_i/G)(R_i/t_\sigma)^2$ expanding with a Lorentz factor $\gamma_{shell}\sim \gamma_i^2 \sim (t_\sigma/t_i)^2$ relative to the Hubble flow.
Possible astrophysical effects of this propagating shell are left as a subject for further research.
 
After the momentum has been transferred to matter, the bubble starts receding relative to the Hubble flow. As we shall see, for large bubbles with $R_i\gg t_i$ the relative shrinking speed becomes relativistic in a time-scale which is much smaller than the Hubble scale. To illustrate this point, let us first consider the case of pressureless matter. 

\subsection{Bubbles surrounded by dust}

\label{bubindust}

Once the bubble wall has stopped with respect to the Hubble flow, it is surrounded by an infinitesimal layer devoid of matter, with a negligible vacuum energy $\rho_0$ which we shall take to be zero. Since the matter outside of this layer cannot affect the motion of the wall, from that time on we can use Eq. (\ref{mass}) with $\rho_i$ replaced by the vanishing vacuum energy after inflation, $\rho_0=0$,
\begin{equation} 
M_{bh}={4\over 3} \pi \rho_b R^3+ 4\pi \sigma R^2 [\dot R^2 +1-H_b^2 R^2]^{1/2} - 8\pi^2 G \sigma^2 R^3. \label{massii}
\end{equation}
The mass parameter $M_{bh}$ for the Schwarzschild metric within the empty layer can be estimated from the initial condition 
$\dot R_i \approx H_i R_i$, at the time $t=t_i+\Delta t \approx t_i$ when the bubble is at rest with respect to the Hubble flow. Using the separation of scales (\ref{separation}), we have 
\begin{equation}
M_{bh}\approx \left({4\over 3} \pi \rho_b + 4\pi\sigma H_i\right) R_i^3. \label{bhmassbubble}
\end{equation}
The motion of the bubble is such that it eventually forms a black hole of mass $M_{bh}$, nested in an empty cavity of co-moving radius 
\begin{equation}
r_c \approx {R_i\over a(t_i)}.
\end{equation}
A remarkable feature of (\ref{bhmassbubble}) is that $M_{bh}$ is proportional to the initial volume occupied by the bubble, with a universal proportionality factor which depends only on microphysical parameters. Under the condition (\ref{separation}), this factor is much smaller than the initial matter density $\rho_m(t_i)$, and we have
\begin{equation}
M_{bh} \sim \left({t_i^2\over t_b^2} + {t_i\over t_\sigma}\right) M_i \ll M_i,\label{bhmassestimate}
\end{equation}
where $M_i$ is the excluded mass given in (\ref{excluded}). On dimensional grounds, the first term in parenthesis in (\ref{bhmassbubble}) is of order $\eta_b^4$, where $\eta_b$ is the energy scale in the vacuum inside the bubble, whereas the second term is of order $\eta_w^3 \eta_i^2/M_p$, where $\eta_w$ is the energy scale of the bubble wall and $\eta_i$ is the inflationary energy scale. The second term is Planck suppressed, and so in the absence of a large hierarchy between $\eta_w$, $\eta_i$ and $\eta_b$, the first term will be dominant. In this case we have $M_{bh}\approx (t_i/t_b)^2 M_i$. However, we may also consider the limit when the vacuum energy inside the bubble is completely negligible, and then we 
have $M_{bh} \sim (t_i/t_\sigma) M_i$.

\begin{figure}
 \includegraphics[width=.7\textwidth]{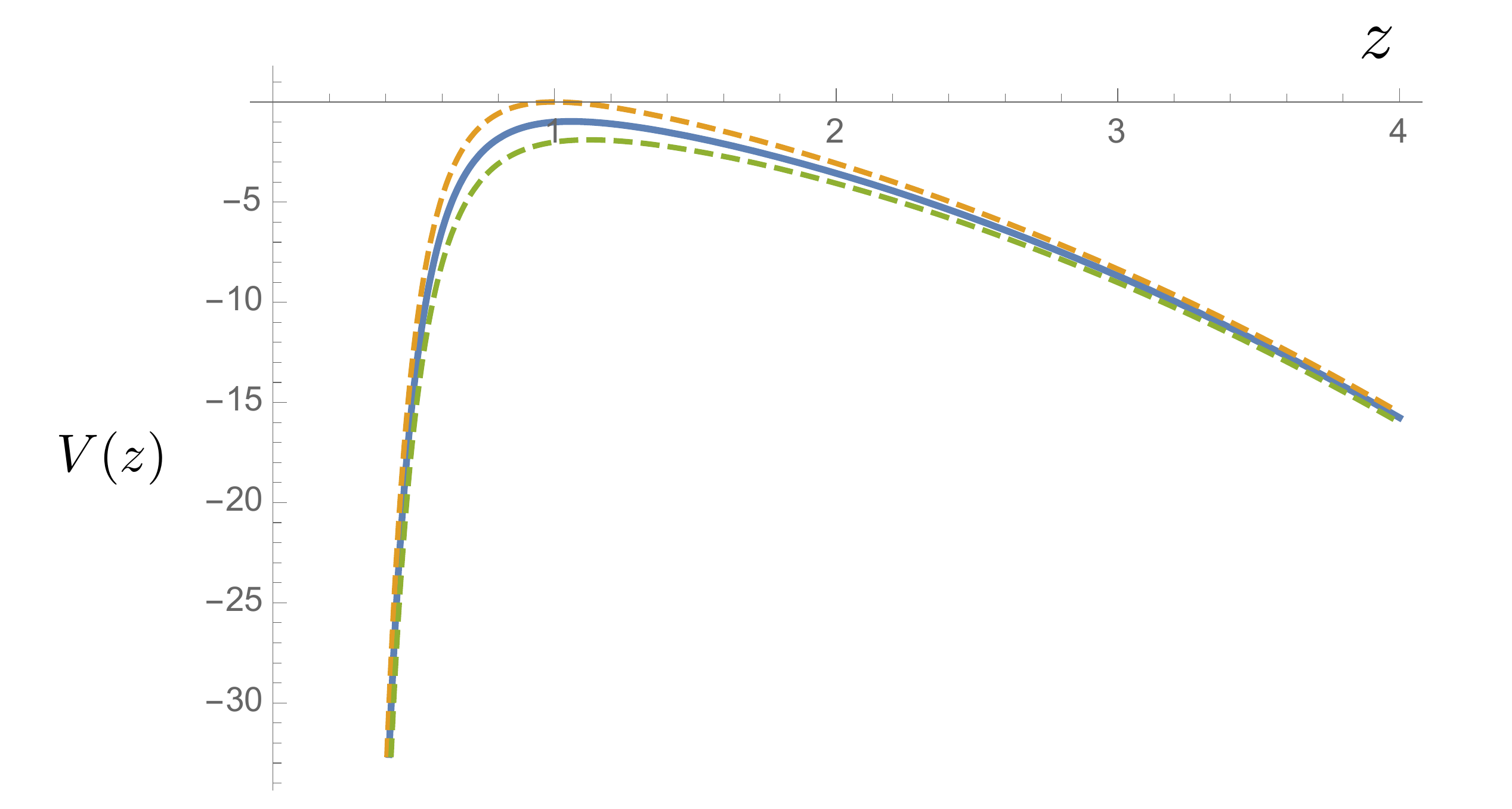}
\caption{The potential in Eq. (\ref{vz}), for $g=0$ (upper curve), $g=1$ (middle curve) and $g=2$ (lower curve).}
\label{vofz}
\end{figure}

It was shown by Blau, Guendelman and Guth \cite{Blau} that the equation of motion (\ref{massii}) can be written in the form
\begin{equation}
\left({dz\over d\tilde\tau}\right)^2 + V(z) = E, \label{conservation}
\end{equation}
where the rescaled variables $z$ and $\tilde \tau$ are defined by
\begin{eqnarray}
z^3&=&{H_+^2 \over 2GM_{bh}} R^3,\\
\tilde\tau&=& {H_+^2\over 4 H_\sigma} \tau,
\end{eqnarray}
with $\tau$ the proper time on the bubble wall trajectory, and $H_+$ defined by
\begin{equation}
H_+^2= H_b^2+4H_\sigma^2.
\end{equation}
The potential and the conserved energy in Eq. (\ref{conservation}) are given by
\begin{eqnarray}
V(z)&=& -\left({1-z^3\over z^2}\right)^2 -{g^2\over z},\label{vz}\\
E&=&{-16 H_\sigma^2 \over (2GM_{bh})^{2/3} H_+^{8/3}},
\end{eqnarray}
where
\begin{equation}
g= {4H_\sigma\over H_+}.
\end{equation}
The shape of the potential is plotted in Fig. \ref{vofz}. The maximum of the potential is at 
\begin{equation}
z^3=z^3_m \equiv {1\over 2}\left[\sqrt{8+(1-g^2/2)^2}-(1-g^2/2)\right].
\end{equation}
Note that $0\leq g\leq 2$, and $1\leq z_m \sim 1$ in the whole parameter range.

The qualitative behaviour of the bubble motion depends on whether the Schwarzshild mass $M_{bh}$ is larger or smaller than a critical mass
defined by
\begin{equation}
M_{cr} = \bar M {g^3 z_m^6(1-g^2/4)^{1/2}\over 3\sqrt{3}(z_m^6-1)^{3/2}},
\end{equation}
where
\begin{equation}
\bar M \equiv {t_b\over 2G}.
\end{equation}
Although the expression for the critical mass is somewhat cumbersome, we can estimate it as
\begin{equation}
GM_{cr} \sim {\rm Min}\{t_\sigma,t_b\}.
\end{equation}
On dimensional grounds $t_b\sim M_p/\eta_b^2$, while $t_\sigma\sim M_p^2/\eta_\sigma^3$. In the absence of a hierarchy between $\eta_b$ and $\eta_\sigma$ 
we may expect $GM_{cr} \sim t_b$, but the situation where $GM_{cr} \sim t_\sigma$ is of course also possible.

\begin{figure}
 \includegraphics[width=.7\textwidth]{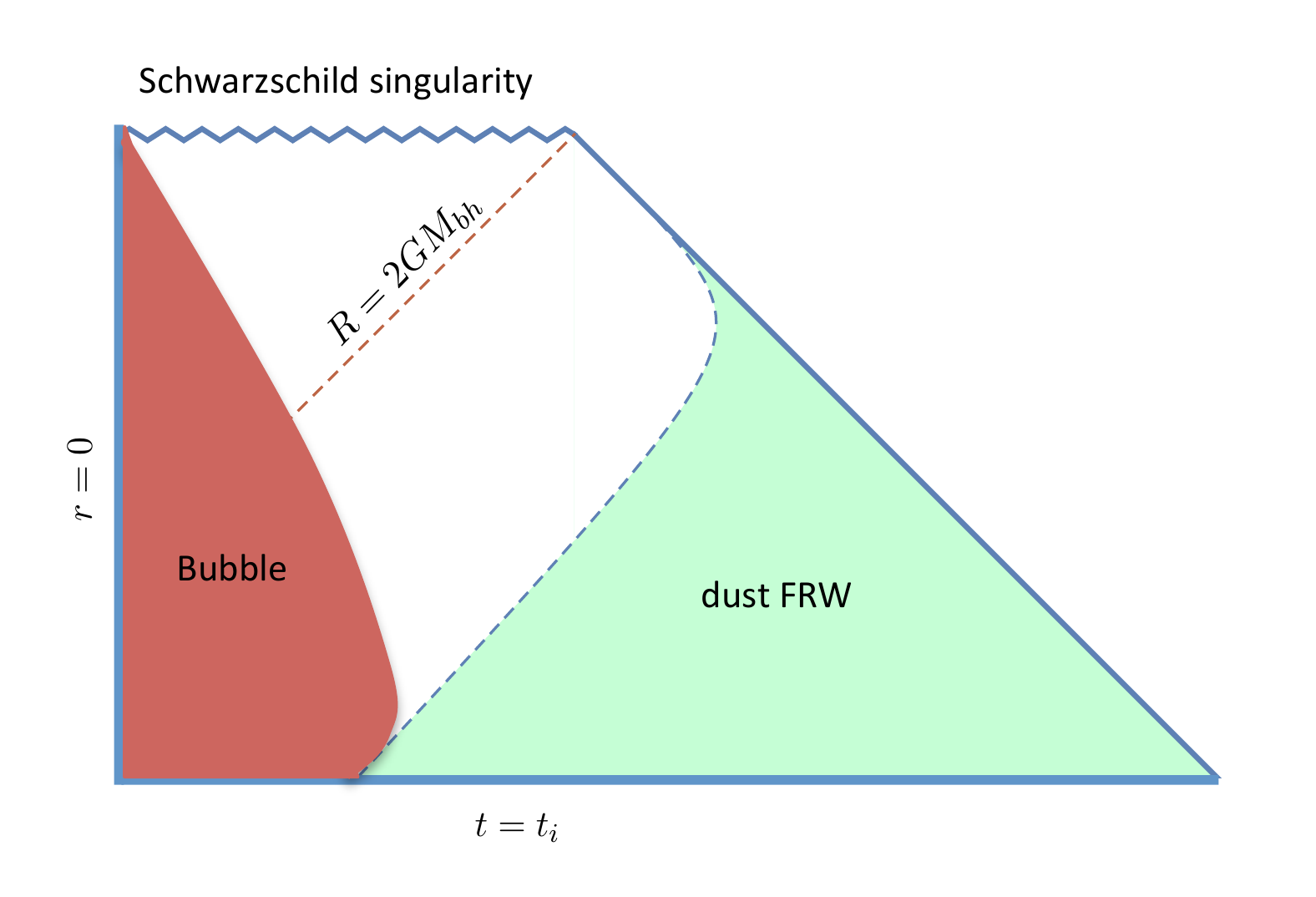}
\caption{Causal diagram showing the formation of a black hole by a small vacuum bubble, with mass $M_{bh}\ll M_{cr}$, in a dust dominated spatially flat FRW.
At the time $t_i$ when inflation ends, a small bubble (to the left of the diagram) with positive energy density $\rho_b>0$, initially expands with a large Lorentz factor relative to the Hubble flow. The motion is slowed down by momentum transfer to matter, turning around due to the internal negative pressure and the tension of the bubble wall, and eventually collapsing into a Schwarzschild singularity. 
At fixed angular coordinates, the time-like dashed line at the edge of the dust region represents a radial geodesic of the Schwarzschild metric (see Appendix \ref{appendixglobal}). Such geodesic approaches time-like infinity with unit slope. The reason is that a smaller slope would correspond to the trajectory of a particle at a finite radial coordinate $R$, while the geodesic of a particle escaping to infinity has unbounded $R$. The thin relativistic expanding shell of matter produced by the bubble is not represented in the figure. This shell disturbs the homogeneity of dust near the boundary represented by the dashed line.
}
\label{smallbh}
\end{figure} 


\subsubsection{Small bubbles surrounded by dust}

For $M_{bh}<M_{cr}$, we have $V(z_m)>E$, and the bubble trajectory has a turning point. In this case it is easy to show that with the condition (\ref{separation}), the bubble trajectory is confined to the left of the potential barrier, $z_i<1\leq z_m$,  where $z_i$ is the value of $z$ which corresponds to the initial radius $R_i$. The bubbles grow from small size up to a maximum radius $R_{max}$ and then recollapse.
The turning point is determined from (\ref{massii}) with $\dot R=0$.  For $M_{bh} \ll M_{cr}$ we have
\begin{equation} 
M_{bh}\approx {4\over 3} \pi \rho_b  R_{max}^3+ 4\pi \sigma R_{max}^2. \label{massii2}
\end{equation}
For $\sigma H_i/\rho_b \ll 1$, the first term in (\ref{bhmassbubble}) dominates, and the turning point occurs after a small fractional increase in the bubble radius
\begin{equation}
{\Delta R\over R_i} \approx {\sigma\over \rho_b} H_i \ll 1,
\end{equation}
which happens after a time-scale much shorter than the expansion time, $\Delta t \sim (\sigma/\rho_b) \ll t_i$. For $1\ll \sigma H_i/\rho_b \ll (R_i H_i)^{3/2}$
we have 
\begin{equation}
{R_{max}\over R_i} \approx \left({3\sigma H_i\over \rho_b}\right)^{1/3}.
\end{equation}
Finally, for $\sigma H_i/\rho_b \gg (R_i H_i)^{3/2}$ we have 
\begin{equation}
{R_{max} \over R_i} \approx (H_i R_i)^{1/2}.\label{ddd}
\end{equation}
In the last two cases, $R_{max} \gg R_i$.  From Eq. (\ref{massii}), we find that for $R_i\lesssim R\ll R_{max}$ the bubble radius behaves as $R\propto \tau^{1/3}$. This is in contrast with the behaviour of the scale factor in the matter dominated universe, $a(t)\propto t^{2/3}$, and therefore the bubble wall decouples from the Hubble flow in a time-scale which is at most of order $t_i$.
Once the bubble radius reaches its turning point at $R=R_{max}$, the bubble collapses into a Schwarzschild singularity. This is represented in Fig. \ref{smallbh}.
Denoting by $t_H$ the time it takes for the empty cavity of co-moving radius $r_c=R_i/a(t_i)$ to cross the horizon, we have 
\begin{equation}
t_H \sim (H_i R_i)^2 R_i.
\end{equation} 
Note that $t_H \gg (H_i R_i)^{1/2} R_i \gtrsim R_{max}$, so the black hole forms on a time-scale much shorter than $t_H$.

\subsubsection{Large bubbles surrounded by dust}

\begin{figure}
 \includegraphics[width=.7\textwidth]{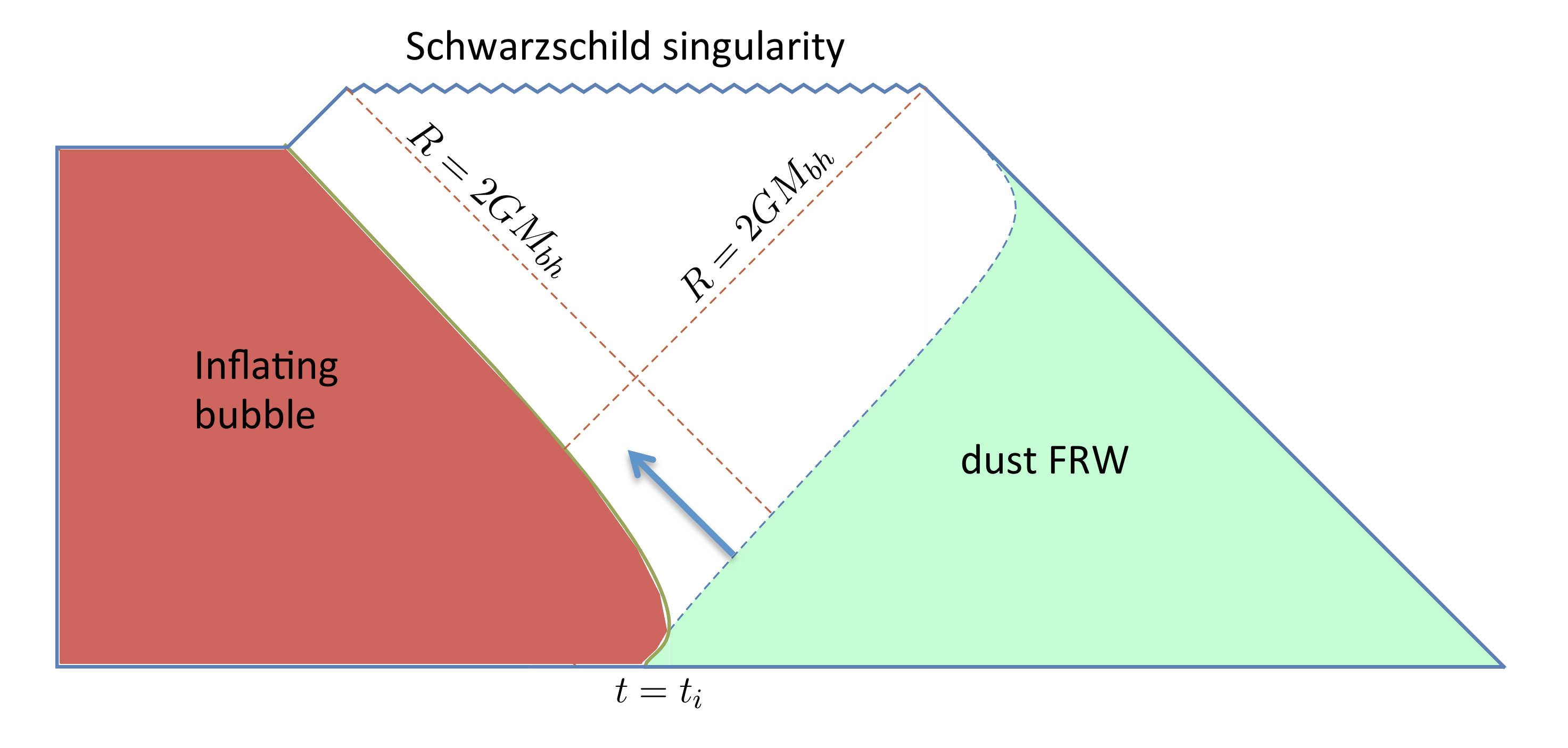}
\caption{Formation of a black hole by a large vacuum bubble, with $M_{bh}\gg M_{cr}$ , in a dust dominated spatially flat FRW. In this case, the bubble does not 
collapse into the Schwarzschild singularity. Instead, at the time $t_{cr} \sim {\rm Min}\{t_b,t_\sigma\}$ the size of the bubble starts growing exponentially in a baby universe, which is connected by a wormhole to the parent dust dominated FRW universe. Initially, a geodesic observer at the edge of the dust region can send signals through the wormhole into the baby universe. This is represented by the blue arrow in the Figure. However, after a proper time $t\sim 2GM_{bh}$, the wormhole ``closes" and any signals which are sent radially inwards end up at the Schwarzschild singularity.}
\label{babymultiverse}
\end{figure}

\begin{figure}
 \includegraphics[width=.5\textwidth]{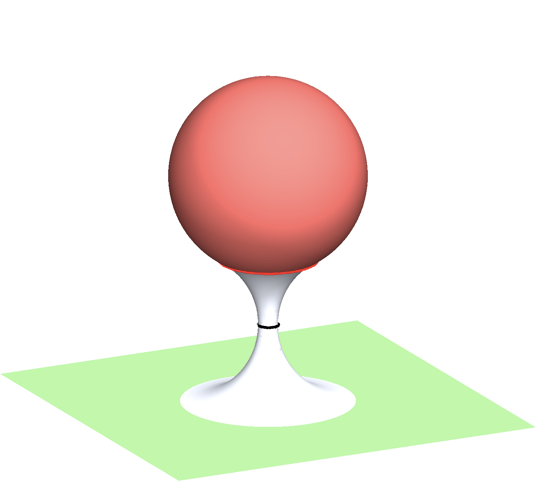}
\caption{A space-like slice of an inflating bubble connected by a wormhole to a dust dominated flat FRW universe.}
\label{vb}
\end{figure}

For $M_{bh}>M_{cr}$ we have $V(z_m)>E$, so the trajectory is  unbounded and the bubble wall grows monotonically towards infinite size in the asymptotic future.  
The unbounded growth of the bubble is exponential, and starts at the time $t \sim {\rm Min}\{t_b,t_\sigma\}$. 
This leads to the formation of a wormhole which eventually ``pinches off", leading to a baby universe (see Figs. \ref{babymultiverse} and \ref{vb}). The bubble inflates at the rate $H_b$,
and transitions to the higher energy inflatinary vacuum with expansion rate $H_i$ will eventually occur, leading to a multiverse structure \cite{LeeWeinberg,recycling}.
Initially, a geodesic observer at the edge of the matter dominated region can send signals that will end up 
in the baby universe. However, after a time $t \sim GM_{bh}$, the wormhole closes and any signals which are sent radially inwards end up at the Schwarzschild singularity.

\subsection{Bubbles surrounded by radiation} \label{bsbr}

 Due to the effects of pressure, the dynamics of bubbles surrounded by radiation is somewhat more involved than 
 in the case of pressureless dust. Here, we will only present a qualitative analysis of this case, since a detailed description will require further numerical studies which are beyond the scope of the present work.

Initially, as mentioned in Subsection \ref{subinitial}, the Lorentz factor of the bubble wall relative to the Hubble flow at the end of inflation is of order $\gamma_i \sim t_\sigma/t_i \gg 1$. As the wall advances into ambient radiation with energy density $\rho_m$, a shock wave forms. 
The shocked fluid is initially at rest in the frame of the wall, and its proper density is of order $\rho_s \sim \gamma^2\rho_m$, where $\gamma$ is the Lorentz factor of the shocked fluid relative to the ambient  \cite{BM}. In the frame of shocked radiation, the shock front moves outward at a subsonic speed $\beta\approx 1/3$, and therefore the
pressure of the shocked gas has time to equilibrate while the wall is pumping energy into it.
Consider a small region of size 
\begin{equation}
L \ll t_i
\end{equation}
near the wall. At such scales, the wall looks approximately flat, and 
gravitational effects (including the expansion of the universe) can be neglected. In the frame of ambient radiation, energy conservation per unit surface of the bubble wall can be expressed as
\begin{equation}
\gamma_i \sigma \sim \gamma^2 \rho_m \Delta t + \gamma  \sigma. \label{ecr}
\end{equation}
The left hand side of this relation represents the initial energy of the wall. The first term in the right hand side is the energy of the layer of shocked fluid. This is
of order $\gamma\rho_s \Delta \tau \sim \gamma^2 \rho_m \Delta t$, where $\Delta \tau$ is the proper time ellapsed since the shock starts forming, and 
$\Delta t=(t-t_i)=\gamma\Delta\tau$ is the corresponding interval in the ambient frame. The second term in the right hand side is just the energy of the wall at time $t$.
Here, we have neglected the rest energy of the ambient fluid, which is of order $\gamma^{-2}$ relative to the shocked fluid.

Eq. (\ref{ecr}) shows that the bubble wall looses most of its energy on a time scale $\Delta t \sim \sigma/(\gamma_i\rho_m) \sim (t_i/t_\sigma)^2 t_i \ll t_i$. 
This is the characteristic time which is required for the pressure of the shocked fluid to accelerate the wall away from it, and is actually in agreement with the rough estimate (\ref{timescale}), which is based on a simplified fluid model. The recoil velocity of the wall cannot grow too large, though, because the backward pressure wave can only travel at the speed of sound. Hence, the relative Lorentz factor $\gamma_w$ between the wall and the shocked fluid will be at most of  order one.
The initial energy of the wall per unit surface is is $\gamma_i \sigma_i \sim G/t_i$. After time of order $t_i$, this energy will be distributed within a layer of width $t_i$ near the position of the wall, and its density will be comparable to the ambient energy density $\rho_m \sim G/t_i^2$. By that time, the shock will have dissipated, and the wall will have, at most, a mildly relativistic speed with respect to the ambient Hubble flow.

For $t\gtrsim t_i$, the expansion of the universe, and of the bubble, start playing a role. The wall's motion becomes then dominated by the effect of its tension, and of the vacuum energy density in its interior. Again, the fate of gravitational collapse depends on whether the bubble is larger or smaller than a critical size.

\subsubsection{Small bubbles surrounded by radiation}

Shortly after $t \sim t_i$, the bubble wall will be recoiling from ambient radiation at mildly relativistic speed, due to the pressure wave which proceeds inward at the speed of sound, in the aftermath of the shock. For $t\gtrsim t_i$, we expect that the wall will only be in contact with rarified radiation  dripping ahead of the pressure wave. Neglecting the possible effect of such radiation, the dynamics proceeds as if the bubble were in an empty cavity, along the lines of our discussion in Subsection \ref{bubindust}. 
In that case, the bubble has a conserved energy $M_{bh}$ given approximately by Eq. (\ref{bhmassbubble}).
For $M_{bh}\ll M_{cr}$, the bubble initially grows until it reaches a turning point, at some maximum radius $R_{max}$, after which it collapses to
form a black hole of mass $M_{bh}$. As mentioned after Eq. (\ref{ddd}), for $R_i\lesssim R \ll R_{max}$, the radius grows as
\begin{equation}
R\propto \tau^{1/3},
\end{equation}
where $\tau$ is proper time on the wall. Note that the ambient radiation of the FRW universe expands faster than that, as a function of its own proper time, since $a(t)\propto t^{1/2}$.
This means that even in the absence of external pressure, the bubble motion decouples from the Hubble flow on the timescale $t_i$, with a tendency for the gap between the bubble and the ambient Hubble flow to grow in time. In what follows, and awaiting confirmation from numerical studies, we assume that the influence of radiation on the wall's motion is negligible thereafter.


Eventually the wall will collapse, driven by the negative pressure in its interior and by the wall tension.
We can estimate the mass of the resulting black hole as 
the sum of bulk and bubble wall contributions, as in Eq. (\ref{bhmassbubble}):
\begin{equation}
M_{bh} \sim \left({4\over 3} \pi \rho_b + 4\pi\sigma H_i \right) R_i^3.\label{bhmassrad}
\end{equation}
In a radiation dominated universe, the co-moving region of initial size $R_i$ crosses the horizon 
at the time
\begin{equation}
t_H \sim {R_i^2 \over t_i},
\end{equation}
and we can write (\ref{bhmassrad}) as
\begin{equation}
GM_{bh} \sim \left({t_i R_i \over t_b^2}  + {R_i\over t_\sigma} \right) t_H \ll t_H, \label{msth}
\end{equation}
where in the last relation we have used that for $M_{bh} \ll M_{cr}$, we have $t_i\lesssim R_i \ll t_b, t_\sigma$. Hence, the black hole radius is much smaller than the 
co-moving region affected by the bubble at the time of horizon crossing.

\subsubsection{Large bubbles surrounded by radiation}

For $M_{bh} \gg M_{cr}$, the situation is more complicated. A wormhole starts developing at $t\sim GM_{cr}$, when the growth of the bubble becomes exponential. On the other hand, the wormhole stays open during a time-scale of order $\Delta t\sim GM_{bh} \gg GM_{cr}$.  During that time, some radiation can flow together with the bubble into the baby universe, thus changing the estimate (\ref{bhmassrad}) for $M_{bh}$. 
On general grounds, we expect that the radius of the black hole is at most of the size of the cosmological horizon at the time $t_H$ when the co-moving 
scale corresponding to $R_i$ crosses the horizon
\begin{equation}
GM_{bh} \lesssim t_H.
\end{equation}
The study of modifications to Eq. (\ref{bhmassrad}) for $M_{bh} \gg M_{cr}$ requires numerical analysis and is left for further research.

\section{Gravitational collapse of domain walls}\label{gcdw}

In the previous sections we considered bubbles with $\rho_b\ll \rho_i$. This condition led to a highly boosted bubble wall at the time when inflation ends. Here we shall concentrate instead on domain walls, which correspond to the limit $\rho_b=\rho_i$. Domain walls will be essentially at rest with respect to the Hubble flow at the end of inflation. Another difference is that for the case of bubbles we assumed that the inflaton field lives only outside the bubble, so its energy is transfered to matter that lives also outside. Here, we will explore the alternative possibility where matter is created both inside and outside the domain wall. This seems to be a natural choice, since domain walls are supposed to be the result of spontaneous symmetry breaking, and we may expect the physics of both domains to be essentially the same. As we shall see, the presence of matter inside the domain wall has a dramatic effect on its dynamics.

For $t\ll t_\sigma$, the graviational field of the walls is completely negligible. Walls of superhorizon size $R\gg t_i$ are initially at rest with respect to the Hubble flow, and are conformally stretched by the expansion of the universe, $R\propto a$. Depending on their initial size, their fate will be different. As we shall see, small walls enter the horizon at a time $t_H\ll t_\sigma$ and then they begin to decouple from the Hubble flow. 
Depending on the interaction of matter with the domain wall, they may either collapse to a black hole singularity, or they may develop long lived remnants in the form of pressure supported bags of matter, bounded by the wall. Larger walls with $t_H \gg t_\sigma$ start creating wormholes at the time $t\sim t_\sigma$, after which the walls undergo exponential expansion in baby universes.

\subsection{Domain walls surrounded by dust}

The case of walls surrounded by dust is particularly simple, since the effect of matter outside the wall can be completely ignored. As we did for bubbles, 
here we may distinguish between small walls, for which self-gravity is negligible, and large walls, for which it is important.

\subsubsection{Small walls}

Small walls are frozen in with the expansion until they cross the horizon, at a time $t_H\ll t_\sigma$. The fate of the wall after horizon crossing depends crucially on the nature of its interaction with matter, which is highly model dependent \cite{Allen}. To illustrate this point we may consider two contrasting limits.

Let us first consider a situation where matter is reflected off the wall with very high probability. In this case, even if the wall tension exerts an inward pull,  matter inside of it keeps pushing outwards.
Initially, this matter has the escape velocity. In a Newtonian description, its positive kinetic energy exactly balances the negative Newtonian potential $V(R)$. However, part of the kinetic energy will be invested in increasing the surface of the domain wall, up to a maximum size $R_{max}$, which can be estimated through the relation
\begin{equation}
\sigma R_{max}^2 \sim -V(R_{max}) \sim {GM_{bag}^2\over R_{max}} \sim {t_H^2\over GR_{max}}. \label{relationrmax}
\end{equation}
Here $M_{bag}\sim t_H/G$ is the mass of the mater contained inside the wall, and we have neglected a small initial contribution of the domain wall to the energy budget. From (\ref{relationrmax}) we have
\begin{equation}
R_{max} \sim (t_H^2 t_\sigma)^{1/3} \gg t_H.
\end{equation}
At that radius the gravitational potential of the system $\Phi\sim V/M$ is very small 
\begin{equation}
-\Phi \sim  (t_H/t_\sigma)^{1/3} \ll 1. \label{gpot}
\end{equation}
After reaching the radius $R_{max}$, the ball of matter inside the wall starts collapsing under its own weight, helped also by the force exerted by the wall tension.  As a result, matter develops a pressure of order $p\sim \rho_m v^2$, where $v^2 \sim GM_{bag}/R$ is the mean squared velocity of matter particles.  The collapse will halt at a radius $R=R_{bag}$ where this pressure balances the wall tension:
\begin{equation}
{\sigma\over R_{bag}} \sim p \sim {GM_{bag}^2 \over R_{bag}^4}.\label{relrbag}
\end{equation}
The relation (\ref{relrbag}) is satisfied for a radius which is comparable to the maximum radius, 
\begin{equation}
R_{bag} \sim R_{max},
\end{equation}
and much larger than the gravitational radius,
\begin{equation}
GM_{bag} \sim t_H \ll R_{bag}.\label{massbagdust}
\end{equation}
The velocity of matter inside the bag is of the order of the virial velocity, and we do not expect any substructures to develop by gravitational instability inside the bag.

Next, let us consider the case of a ``permeable'' wall, such that matter can freely flow from one side of it to the other. For $t_H\ll t_\sigma$, this will simply behave as a test domain wall in an expanding FRW. Once the wall falls within the horizon, it will shrink under its tension much like it would in flat space, and the mass of the black hole can be estimated as
\begin{equation}
M_{bh} \approx 9\pi C_{m}\, \sigma t_H^2. \label{small1}
\end{equation}
The coefficient $C_{m}$ can be found by means of a numerical study. This is done in Appendix A.

\begin{figure}
 \includegraphics[width=.7\textwidth]{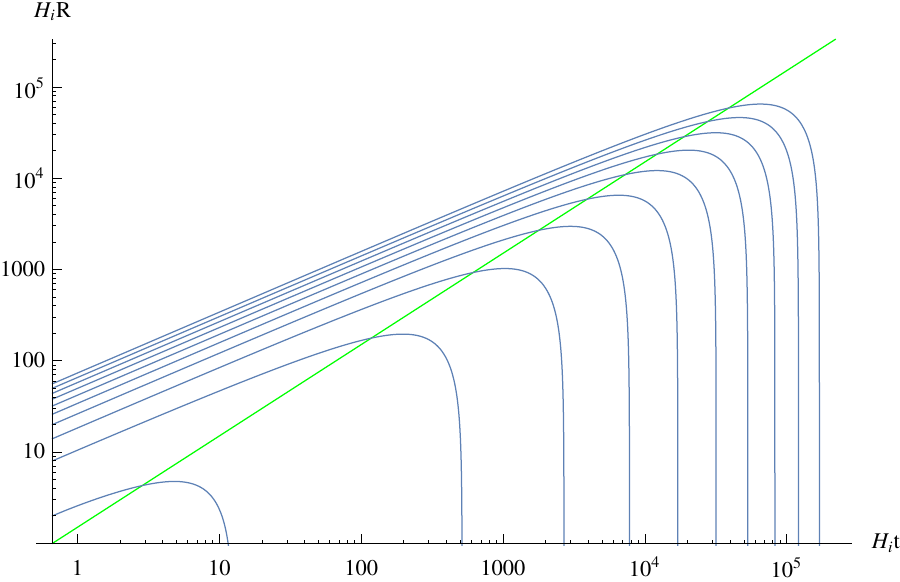}
\caption{Evolution of the radius of a test domain wall in a matter dominated universe, as a function of cosmological time, for different initial radii. The green line
corresponds to the horizon crossing time $t_H$, when $R=H^{-1}$.}
\label{rt}
\end{figure}

A test wall can be described by using the Nambu action in an expanding universe. The mass of the wall is defined as
\begin{equation}
M(t) = 4\pi \sigma {a^2 r^2 \over \sqrt{1-a^2r'^2}},
\label{mwall}
\end{equation}
where the comoving radial coordinate $r(t)$ satisfies the differential equation
\begin{equation}
r''+ (4-3a^2 {r'}^2) H {r'} + \frac{2}{r a^2}(1-a^2{r'}^2)=0.
\end{equation}
Here a prime denotes derivative with respect to cosmic time $t$.  The condition $r'(t_i)=0$ is imposed at some initial 
time such that $R(t_i) \gg t_i$. Fig. \ref{rt} shows the evolution of the domain wall physical radius $R=ar$ as a function of time.

We find that while the wall is stretched by the expansion the mass $M(t)$ is increasing.
Defining $t_H$ through the relation
\begin{equation}
R(t_H)= H^{-1}= (3/2) t_H,
\end{equation}
the wall starts recollapsing around the time $t_H$ and the mass $M(t)$ approaches the constant value 
given by Eq. (\ref{small1}) once $R\ll t_H$. We find that, in the limit $R_i\gg t_i$, the numerical coefficient approaches  
the value 
\begin{equation}
C_{m} \approx 0.15. \label{cm}
\end{equation}
This result is in agreement with a more sophisticated calculation done in Ref. \cite{Yoo}. We note, however, that the result (\ref{cm}) only applies for sufficiently small walls. 
Such walls collapse at $t_H\lesssim t_\sigma$, and the resulting black holes have mass
\beq
M_{bh}\lesssim M_{cr} \sim 1/G^2 \sigma.
\eeq
For larger walls, self-gravity is important, and the scaling $M_{bh} \propto \sigma t_H^2$ no longer holds, as we shall now describe.

\subsubsection{Large domain walls surrounded by dust}\label{largewallindust}

If a domain wall has not crossed the horizon by the time $t_\sigma$,  its radius starts growing exponentially and decouples from the Hubble flow,  forming a wormhole.

To illustrate this process, we may consider an exact solution where a spherical domain wall of initial radius  
\begin{equation}
R_*>t_\sigma, \label{inrad}
\end{equation}
is surrounded by pressuereless dust. The solution is constructed as follows. At some chosen time $t_*$, we assume that the metric inside and outside the wall is initially a flat FRW metric with scale factor $a\propto t^{2/3}$. 
We shall also assume that the wall is separated from matter, on both
sides, by empty layers of infinitesimal width.  (As we shall see, these
layers will grow with time.)
By symmetry, the metric within these two empty layers takes the Schwarzschild form 
\begin{equation}
ds^2=-\left(1-{2GM^\pm\over R}\right) dT^2 + \left(1-{2GM^\pm\over R}\right)^{-1} dR^2 + R^2 d\Omega^2.
\end{equation}
The mass parameter $M^-$ for the interior layer can in principle be different from the mass parameter $M^+$ for the exterior. Let us now find such mass parameters by matching the metric in the empty layers to the adjacent dust dominated FRW. 

\begin{figure}
 \includegraphics[width=1\textwidth]{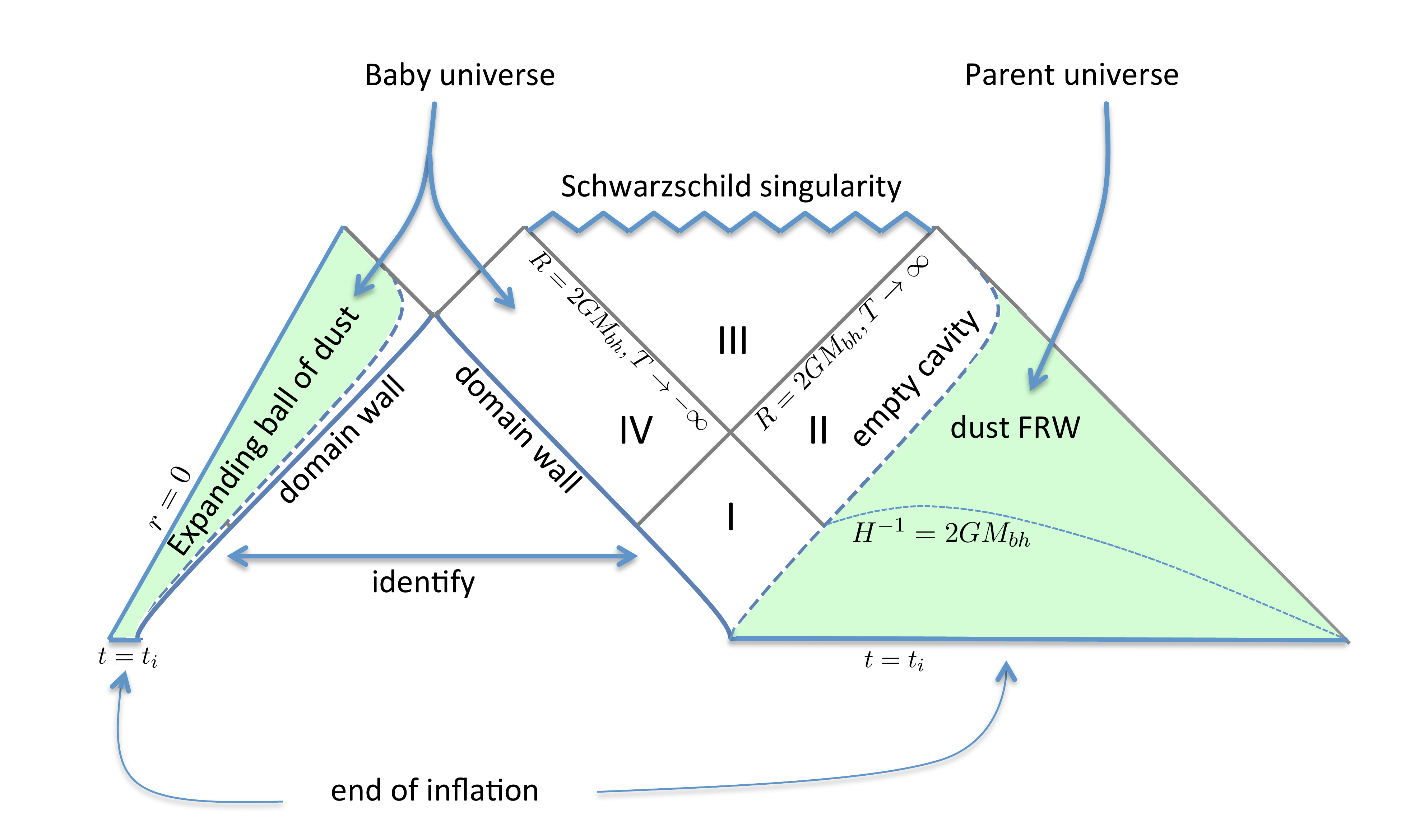}
\caption{A large domain wall causing the birth of a baby universe.
 }
\label{largewall}
\end{figure}

By continuity, the particles of matter at the boundary of a layer must follow a geodesic of both Schwarzschild and FRW. Such geodesics originate at the white-hole/cosmological singularity, and they end at time-like infinity (see Fig. \ref{largewall}). 
The Schwarzschild metric is independent of $T$, and  geodesics satisfiy the conservation of the corresponding canonical momentum:
\begin{equation}
 {\dot T}  =  \left(1 - {2GM^\pm\over R}\right)^{-1}C_\pm.\label{momcons}
\end{equation}
Here $C_\pm$ is a constant, and a dot indicates derivative with respect to proper time. Initially, the geodesic is in the white hole part of the Schwarzschild solution (Region I in Fig.
\ref{largewall}), and $C_\pm$ is positive or negative depending on whether the particle moves to the right or to the left.
Using $(1 - 2GM^\pm/R) {\dot T}^2  -  (1 - 2GM^\pm/R)^{-1} {\dot R}^2 = 1$, we have 
\begin{equation}
{\dot R}^2 = {2GM^\pm\over R} + C_\pm^2 - 1. \label{econs}
\end{equation}
For geodesics with the escape velocity, we have 
\footnote{We conventionally choose the constants $C_\pm$ to be positive 
so that the geodesics originating at the white hole singularity move to the right, from region I towards region II in the extended Schwarzshild diagrams (see Fig. \ref{largewall}). Note that we should think of  the two empty layers on both sides of the wall as segments of two separate Schwarzschild solutions. In Fig. \ref{largewall} these are depicted side by side. The solution for the interior vacuum layer is on the left, and the solution for the exterior vacuum layer is on the right. }
\begin{equation}
C_\pm =1. \label{negativec}
\end{equation}
The equation of motion (\ref{econs}) then takes the form 
\begin{equation}
\left({\dot R\over R}\right)^2 = {2GM^\pm\over R^3}. \label{geofried}
\end{equation}
This has the solution $R\propto t^{2/3}$, where $t$ is proper time along the geodesic of the dust particle, matching the behaviour of the flat FRW expansion factor.
The geodesics at the boundaries of the two infinitesimal empty layers initially coincide at $R=R_*$. Assuming that the initial matter density is the same inside and outside the wall, 
comparison of (\ref{geofried}) with the Friedmann equation leads to the conclusion that $M^+$ and $M^-$ have the same value:
\begin{equation}
M_{bh} \equiv M^{\pm} = {4\pi \over 3}\rho_m(t_*) R^3_*= {3 t_H\over 4G}.
\end{equation}
Here, $M_{bh}$ is simply the total mass of matter inside the domain wall, which is also the mass which is excised from the exterior FRW, and $t_H$ is the time at which 
the boundary of such excised region crosses the horizon.

The dynamics of the bubble wall in between the two empty layers is determined from Israel's matching conditions, which lead to the equations \cite{Blau}
\begin{eqnarray}
\dot T = \pm\left(1-{2GM_{bh}\over R}\right)^{-1}{R\over t_\sigma}, \label{dottwall}
\end{eqnarray}
and
\begin{equation}
\dot R^2 ={2GM_{bh}\over R}+\left({R\over t_\sigma}\right)^2-1. \label{dotrwall}
\end{equation}
The double sign $\pm$ in (\ref{dottwall}) refers to the embedding of the domain wall as seen from the interior and exterior Schwarzshild solutions respectively. 
The embedding is different, since there has to be a jump in the extrinsic curvature of the worldsheet as we go from one side of the wall to the other.
Comparing (\ref{dottwall}) and (\ref{dotrwall}) with (\ref{momcons}) and (\ref{econs})  with $C_\pm=1$ and $M^\pm=M_{bh}$, we see that, for $R>t_\sigma$,
the wall moves faster than the dust particle at the edge of the expanding interior ball of matter. As seen from the outside, the wall moves in the direction opposite to 
the exterior FRW Hubble flow. Consequently, the wall never runs into matter and its motion is well described by (\ref{dotrwall}) thereafter. 

\begin{figure}
 \includegraphics[width=.7\textwidth]{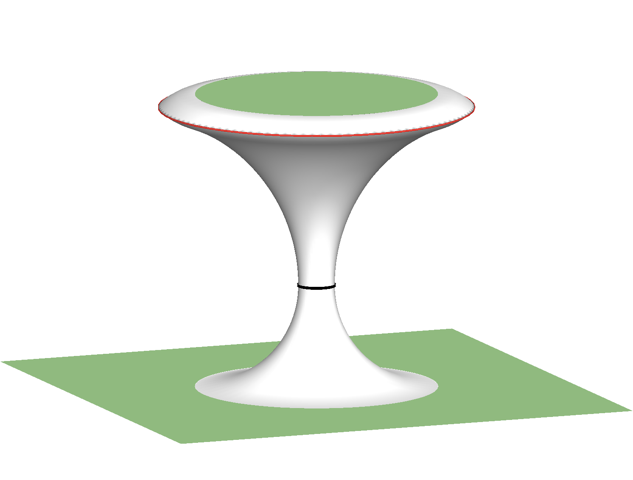}
\caption{A space-like slice of a baby universe with an expanding ball of matter in it, connected by a wormhole to an asymptotic dust dominated FRW universe. The wormhole is created by the repulsive gravitational field
of an inflating domain wall, depicted as a red line.}
\label{dw}
\end{figure}

For $R>t_\sigma$ Eq. (\ref{dotrwall}) has no turning points and the radius continues to expand forever. The expansion reaches exponential 
behaviour $R\propto e^{\tau/t_\sigma}$ after the time when the first term in the right hand side of (\ref{dotrwall}) can be neglected and the second one becomes dominant.
Here $\tau$ is the proper time on the worldsheet.
This expansion is of course much faster than that of the Hubble flow, and takes place in the baby universe (see Figs. \ref{largewall} and \ref{dw}) . Loosely speaking, a wormhole in the extended Schwarzschild solution
forms and closes on a time-scale $\Delta t\sim GM_{bh}$. This coincides with the time $t_H$ when the 
empty cavity containing the black hole crosses the horizon.  An observer at the edge of the cavity can initially send signals into the baby universe, but not after the time when $H^{-1}=2GM_{bh}$.

In the above discussion, we did not make any assumptions about the time $t_*$. The exact solution illustrating the formation of a wormhole is obtained by a matching procedure, and can be constructed for any initial value of $t_*$. As we have seen, the condition $R_*>t_\sigma$ ensures that the wall will not run into matter. The wall evolves in vacuum, and the Hubble flow of
matter remains undisturbed after $t_*$. Moreover, it is easy to show that for $t_*\ll t_\sigma$ and $R_*>t_\sigma$, the wall will be approximately co-moving with the Hubble flow at the time $t_*$, 
with a relative velocity of order $v\sim (t_*/t_\sigma)\ll 1$.\footnote{This can be seen as follows. Denoting by $U^{\mu}$ the four-velocity of matter at the edge of the cavity and by $W^{\mu}$ the four-velocity of the domain wall, 
the relative Lorentz factor is given by
\begin{equation}
\gamma_w =-g_{\mu\nu} U^{\mu}W^{\nu} =\left(1-{9t_*^2 \over 4 R_*^2}\right) \left[\sqrt{1+{9t_*^2\over4 t_\sigma^2}-\left({3 t_*\over 2 R_*}\right)^4} \mp {9t_*^2\over 4 R_*t_\sigma}\right].
\end{equation}
where we have used (\ref{momcons}), (\ref{econs}), (\ref{dottwall}) and (\ref{dotrwall}), with $2GM_{bh}=(4/9t_*^2)R_*^3$.  Using $t_*\ll t_\sigma$ and $R_*>t_\sigma$, we have
$\gamma_w = 1+O(t_*^2/t_\sigma^2)$, or $v\sim (t_*/t_\sigma)\ll 1$.}

Nonetheless, in order to make contact with the setup of our interest, we must consider a broader set of initial conditions such that the wall can be smaller than $t_\sigma$
at the time $t_i$ when inflation ends, $R_i< t_\sigma$. In this case, the wall may initially expand a bit slower than the matter inside.  Consequently some matter may come into contact with the wall. 
However, since superhorizon walls are approximately co-moving for $t\ll t_\sigma$, the effect of this interaction will be small, and we expect the qualitative behaviour to be very similar to that of the exact solution
discussed above provided that the size of the wall becomes larger than $t_\sigma$ at some time $t_*\lesssim t_\sigma$.

\subsection{Walls surrounded by radiation}

\subsubsection{Small walls}

Let us start by discussing the situation where the domain wall is ``impermeable''. Small walls with $t_H\ll t_\sigma$ continue to be dragged by the Hubble flow even after they have crossed the horizon. Unlike dust, radiation outside the wall may continue exerting pressure on it, and initially its effect cannot be neglected. The radius of the wall will stop growing once the wall tension balances the difference in pressure $\Delta p$ between the inside and the outside: 
\begin{equation}
\Delta p \sim {\sigma \over R}. \label{eq}
\end{equation}
After that, the pressure outside decays as $(t_H/t)^2$ and becomes negligible compared with the pressure inside, 
which is given by $p\sim (1/G t_H^2)(t_H/R)^4$. From (\ref{eq}) we then find that the equilibrium radius is of order
\begin{equation}
R_{bag} \sim (t_\sigma/t_H)^{1/3} t_H \gg t_H.
\end{equation}
The mass of the bag of radiation is of order
\begin{equation}
M_{bag} \sim {t_H \over G}\left({t_H\over t_\sigma}\right)^{1/3}.\label{massbagradiation}
\end{equation}
Therefore the gravitational potential at the surface of the bag is very small $\Phi\sim (t_H/t_\sigma)^{2/3}\ll 1$.

Next, we may consider the case of ``permeable" walls. Like in the case of dust, the dynamics of small walls is well described by the Nambu action in a FRW universe. The mass of the resulting black holes can be estimated as
\begin{equation}
M_{bh} \approx 16\pi C_r\, \sigma t_H^2,
\end{equation}
where in a radiation dominated universe $t_H$ is determined from the condition
\begin{equation}
R(t_H) = 2 t_H.
\end{equation}
The numerical coefficient is found by using the numerical approach described in Appendix A, and we find
\begin{equation}
C_r \approx 0.62 .
\end{equation}
Note that, although permeable walls do form black holes, both in the dust dominated and radiation dominated eras, their masses are much smaller than the masses of the corresponding bags of matter trapped by impermeable walls.

\subsubsection{Large domain walls surrounded by radiation}

For $t_H\sim t_\sigma$ the estimates for $M_{bh}$ and $M_{bag}$ which we found in the previous subsection converge to the value 
\begin{equation}
M_{bh} \sim {t_H \over G}\label{radest}.
\end{equation}
For $t_H\gg t_\sigma$ a wormhole will start forming near the time $t_\sigma$, and 
some of the radiation may follow the domain wall into the baby universe. Nonetheless, the size of the resulting black hole embedded in the FRW universe cannot be larger than the cosmological horizon, and therefore $M_{bh} \leq t_H/2G$. Continuity suggests that 
the estimate (\ref{radest}) may be valid also for large bubbles in a radiation dominated universe. A determination of $M_{bh}$ for this case requires a numerical simulation and is left for further research.

\section{Mass distribution of black holes}\label{mdbh}

In the earlier sections we have outlined a number of scenarios whereby black holes can be formed by domain walls or by vacuum bubbles nucleated during inflation.  The resulting mass distribution of black holes depends on the microphysics parameters characterizing the walls and bubbles and on their interaction with matter.  We will not attempt to explore all the possibilities here and will focus instead on one specific case: domain walls which interact very weakly with matter (so that matter particles can freely pass through the walls).  A specific particle physics example could be axionic domain walls, whose couplings to the Standard Model particles are suppressed by the large Peccei-Quinn energy scale.  The walls could also originate from a "shadow" sector of the theory, which couples to the Standard Model only gravitationally.

\subsection{Size distribution of domain walls}

Spherical domain walls nucleate during inflation having radius $R\approx H_i^{-1}$; then they are stretched by the expansion of the universe.  At $t-t_n \gg H_i^{-1}$, where $t_n$ is the nucleation time, the radius of the wall is well approximated by\footnote{As in previous Sections, we assume that $H_i^{-1}\ll t_\sigma$, so wall gravity can be neglected during inflation.}
\beq
R(t) \approx H_i^{-1} \exp [H_i(t-t_n)].
\label{Rt}
\eeq
The wall nucleation rate per Hubble spacetime volume $H_i^{-4}$ is 
\beq
\lambda = A e^{-S} .
\label{lambdawall}
\eeq
In the thin-wall regime, when the wall thickness is small compared to the Hubble radius $H_i^{-1}$, the tunneling action $S$ is given by \cite{Basu}
\beq
S = 2\pi^2 \sigma /H_i^3.
\label{Swall}
\eeq
This estimate is valid as long as $S\gg 1$, or 
\beq
\sigma\gtrsim H_i^3.  
\eeq
With this assumption, the prefactor has been estimated in \cite{Garriga} as 
\beq
A\sim \left(\sigma/H_i^3\right)^2.
\label{A}
\eeq

The number of walls that form in a coordinate interval $d^3{\bf x}$ and in a time interval $dt_n$ is
\beq
dN = \lambda H_i^4 e^{3H_i t_n} d^3{\bf x} dt_n .
\label{dNtn}
\eeq
Using Eq.~(\ref{Rt}) to express $t_n$ in terms of $R$, we find the distribution of domain wall radii \cite{Basu},
\beq
dn\equiv \frac{dN}{dV} = {\lambda} \frac{dR}{R^4} ,
\label{dN}
\eeq
where 
\beq
dV = e^{3H_i t}d^3{\bf x}
\eeq
is the physical volume element.  At the end of inflation, the distribution (\ref{dN}) spans the range of scales from $R\sim H_i^{-1}$ to $R_{max} \sim H_i^{-1}\exp({\cal N}_{inf})$, where ${\cal N}_{inf}$ is the number of inflationary e-foldings.  For models of inflation that solve the horizon and flatness problems, the comoving size of $R_{max}$ is far greater than the present horizon.

After inflation, the walls are initially stretched by the expansion of the universe, 
\beq
R(t)=\frac{a(t)}{a(t_i)} R_i,
\label{Rtti}
\eeq
where $a(t)$ is the scale factor, $t_i$ corresponds to the end of inflation and $R_i=R(t_i)$.  The size distribution of the walls during this period is still given by Eq.~(\ref{dN}).  

\subsection{Black hole mass distribution}

As we discussed in Section III, the ultimate fate of a given domain wall depends on the relative magnitude of two time parameters: the Hubble crossing time $t_H$ when $R(t_H) = H^{-1}$, where $H={\dot a}/a$ is the Hubble parameter, and the time $t_\sigma \sim H_\sigma^{-1}=(2\pi G\sigma)^{-1}$, when the repulsive gravity of the wall becomes dynamically important.  Walls that cross the Hubble radius at $t_H \ll t_\sigma$ have little effect on the surrounding matter and can be treated as test walls in the FLRW background.  The mass of such walls at Hubble crossing (disregarding their kinetic energy) is
\beq
M_H \approx 4\pi \sigma H^{-2} \sim \frac{t_H}{t_\sigma} {\cal M}(t_H) \ll {\cal M}(t_H),
\label{MH}
\eeq
where
\beq
{\cal M}(t) = \frac{1}{2GH}
\eeq
is the mass within a sphere of Hubble radius at time $t$.  Once the wall comes within the Hubble radius, it collapses to a black hole of mass $M = CM_H$ with $C\sim 1$ in about a Hubble time.  

In the opposite limit, $t_H\gg t_\sigma$, the wall starts expanding faster than the background at $t\sim t_\sigma$ and develops a wormhole, which is seen as a black hole from the FRW region.  Assuming that the universe is dominated by nonrelativistic matter, we found that the mass of this black hole is
\beq
M\sim {\cal M}(t_H)
\label{M2}
\eeq
and its Schwarzschild radius is $r_g \sim t_H$.  The boundary between the two regimes, $t_H\sim t_\sigma$, corresponds to black holes of mass $M \sim 1/G^2\sigma \sim M_{cr}$, where $M_{cr}$ is the critical mass introduced in Section III.  We note that depending on the wall tension $\sigma$, the critical mass $M_{cr}$ can take a wide range of values, including values of astrophysical interest.  If we define the energy scale of the wall $\eta$ as $\sigma\sim \eta^3$, then, as $\eta$ varies from $\sim 1~GeV$ to the GUT scale $\sim 10^{16}GeV$, the critical mass varies from $10^{17} M_\odot$ to $10^4 g$.

The estimate (\ref{M2}) applies for $t_H\gtrsim t_\sigma \gtrsim t_{eq}$, where $t_{eq}$ is the time of equal matter and radiation densities.  For walls that start developing wormholes during the radiation era, $t_\sigma < t_{eq}$, we could only obtain an upper bound,
\beq
M \lesssim {\cal M}(t_H) ~~~~~ (t_\sigma < t_{eq}).
\label{Mbound}
\eeq

We shall start with the case $t_\sigma > t_{eq}$, which can be fully described analytically.  
Note, however, that this case requires the energy scale of the wall to be $\eta\lesssim 1~GeV$, which is rather small by particle physics standards.  The critical mass in this case is $M_{cr}>10^{17} M_\odot$.

\subsubsection{$t_\sigma > t_{eq}$}

Let us first consider black holes resulting from domain walls that collapse during the radiation era, $t_H < t_{eq}$, and have radii between $R$ and $R+dR$ at Hubble crossing.  For such domain walls, the Hubble crossing time is $t_H=R/2$, and their density at $t>t_H$ is
\beq
dn(t) \sim \lambda \left(\frac{R}{t}\right)^{3/2} \frac{dR}{R^4}.
\label{dn}
\eeq
The mass distribution of black holes can now be found by expressing $R$ in terms of $M$, $R\sim (M/ \sigma)^{1/2}$, and substituting in Eq.~(\ref{dn}):
\beq
dn = {\lambda} \left(\frac{\sigma}{t^2}\right)^{3/4}\frac{dM}{M^{7/4}}.
\label{dn1}
\eeq
A useful characteristic of this distribution is the mass density of black holes per logarithmic mass interval in units of the dark matter density $\rho_{DM}$,
\beq
f(M) \equiv \frac{M^2}{\rho_{DM}} \frac{dn}{dM}.
\label{beta}
\eeq
Since black hole and matter densities are diluted in the same way, $f(M)$ remains constant in time.  We shall evaluate it at the time of equal matter and radiation densities, $t_{eq}$, when Eq.~(\ref{dn}) derived for the radiation era should still apply by order of magnitude.  With $\rho_{DM}(t_{eq}) \sim B^{-1} G t_{eq}^2$, where $B\sim 10$, we have
\beq 
f(M) \sim \lambda (\sigma)^{3/4} G t_{eq}^{1/2} M^{1/4} \sim B \lambda \frac{{\cal M}_{eq}^{1/2} M^{1/4}}{M_{cr}^{3/4}}. 
\label{beta1}
\eeq
Here, ${\cal M}_{eq}\sim t_{eq}/2G\sim 10^{17} M_\odot$ is the dark matter mass within a Hubble radius at $t_{eq}$.

\begin{figure}
 \includegraphics[width=.7\textwidth]{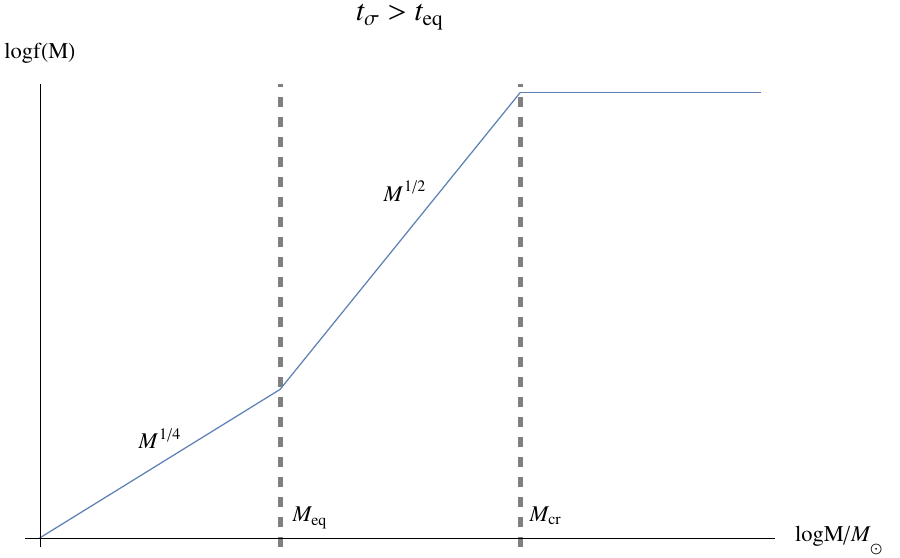}
\caption{Mass distribution function for $t_\sigma>t_{eq}$.}
\label{f1}
\end{figure}

For black holes forming in the matter era, but before $t_\sigma$ ($t_\sigma > t_H > t_{eq}$), Eqs.~(\ref{dn}) and (\ref{beta1}) are replaced by
\beq
dn = \lambda\left(\frac{R}{t}\right)^2 \frac{dR}{R^4},
\label{dn3}
\eeq
\beq
f(M) \sim B\lambda\left(\frac{M}{M_{cr}}\right)^{1/2}.
\eeq
Finally, for black holes formed at $t > t_\sigma$, the mass is 
\beq
M\sim {\cal M} (t_H) \sim R/G.
\eeq
Substituting this in Eq.~(\ref{dn3}), we find 
\beq
f(M)\sim B\lambda.
\eeq
The resulting mass distribution function is plotted in Fig. \ref{f1}.

\subsubsection{$t_\sigma < t_{eq}$}

We now turn to the more interesting case of $t_\sigma < t_{eq}$.  (We shall see that observational constraints on our model parameters come only from this regime.) In this case, domain walls start developing wormholes during the radiation era, and we have only an upper bound (\ref{Mbound}) for the mass of the resulting black holes.  

\begin{figure}
 \includegraphics[width=.7\textwidth]{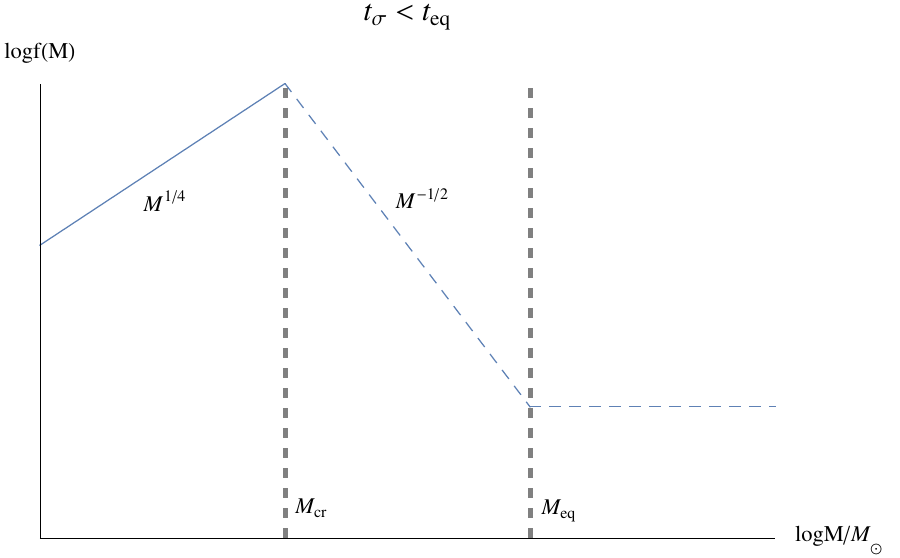}
\caption{Mass distribution function for $t_\sigma<t_{eq}$.}
\label{f2}
\end{figure}

Domain walls that have radii $R < t_\sigma$ at horizon crossing collapse to black holes with $M < M_{cr}$.  The size distribution of such walls and the mass distribution of the black holes are still given by Eqs.~(\ref{dn1}) and (\ref{beta1}), respectively.  For $t_\sigma < R < t_{eq}$, let us assume for the moment that the bound (\ref{Mbound}) is saturated.  Then, using $R\sim GM$ in Eq.~(\ref{dn}), we obtain the mass distribution
\beq
f(M) \sim B \lambda \left(\frac{{\cal M}_{eq}}{M} \right)^{1/2},
\label{beta2}
\eeq
where $B\sim 10$, as before.  This distribution applies for $M_{cr}<M<{\cal M}_{eq}$.  With the same assumption, for walls with $R>t_{eq}$ we find
\beq
f(M)\sim B \lambda.
\label{beta4}
\eeq
The resulting mass distribution function is plotted in Fig. \ref{f2}, with the parts depending on the  assumed saturation of the mass bound shown by dashed lines.

The assumption that the bound (\ref{Mbound}) is saturated for $M>M_{cr}$ appears to be a reasonable guess.  We know that it is indeed saturated in a matter-dominated universe, and it yields a mass distribution that joins smoothly with the distribution we found for $M < M_{cr}$.  A reliable calculation of $f(M)$ in this case should await numerical simulations of supercritical domain walls in a radiation-dominated universe.  For the time being, the distribution we found here provides an upper bound for the black hole mass function.

\section{Observational bounds}\label{ob}

Observational bounds on the mass spectrum of primordial black holes have been extensively studied; for an up to date review, see, e.g., \cite{Carr}.  For small black holes, the most stringent bound comes from the $\gamma$-ray background resulting from black hole evaporation:
\beq
f(M\sim 10^{15}~g) \lesssim 10^{-8}.
\label{Mbound1}
\eeq
For massive black holes with $M > 10^3 M_\odot$, the strongest bound is due to distortions of the CMB spectrum produced by the radiation emitted by gas accreted onto the black holes \cite{Ostriker}:
\beq
f(M> 10^{3}~M_\odot) \lesssim 10^{-6}.
\label{Mbound2}
\eeq  
Of course, the total mass density of black holes cannot exceed the density of the dark matter.  Since the mass distribution in Fig. \ref{f2} is peaked at $M_{bh}=M_{cr}$, this implies
\beq
f(M_{cr})<1.
\label{Mbound3}
\eeq
These bounds can now be used to impose constraints on the domain wall model that we analyzed in Section IV.
As before, we shall proceed under the assumption that the mass bound (\ref{Mbound}) is saturated.  

\begin{figure}
 \includegraphics[width=0.6\textwidth]{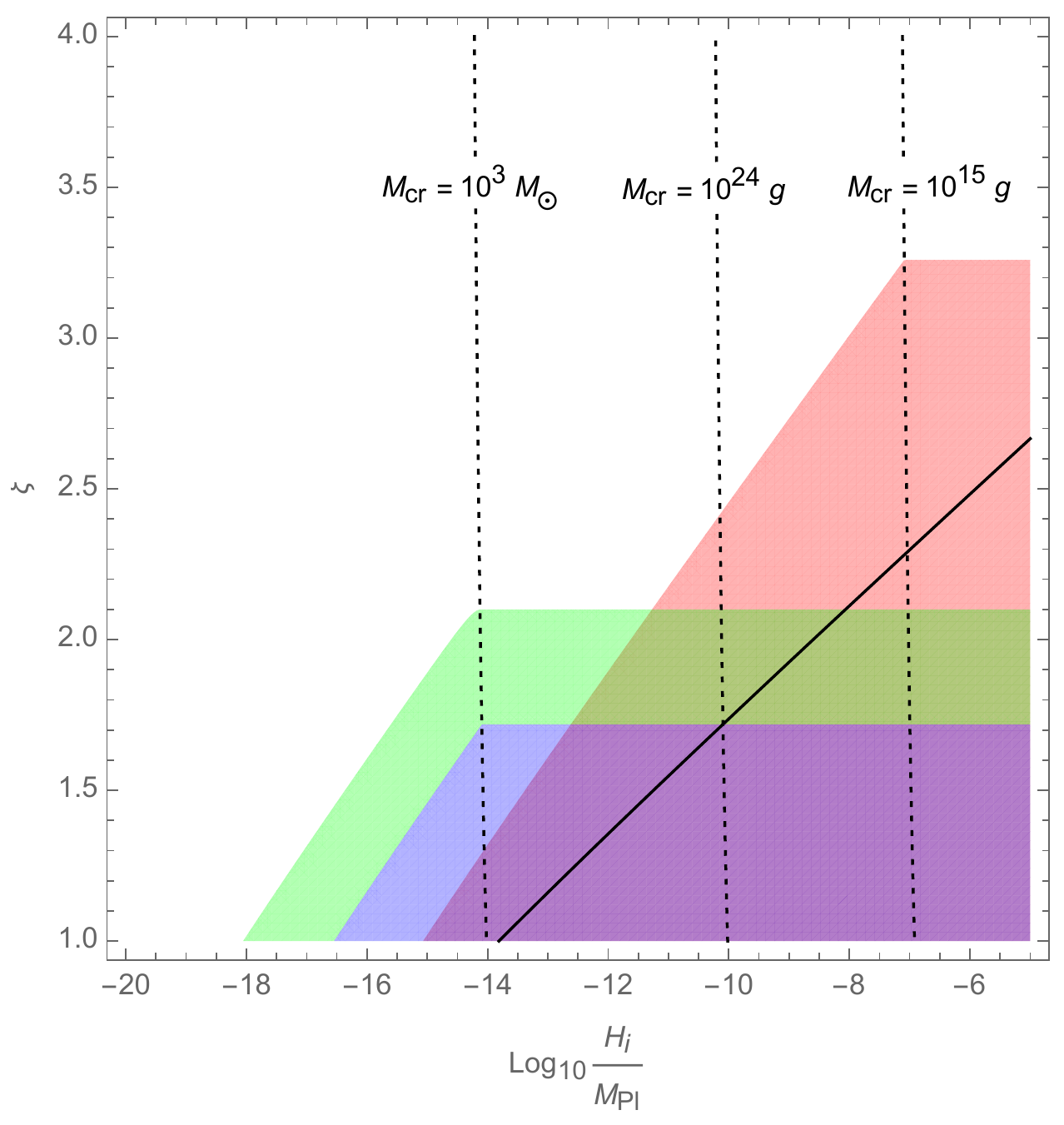}
\caption{Observational constraints on the distribution of black holes produced by domain walls.  Red and purple regions mark the parameter values excluded, respectively,  by small black hole evaporation and by gas accretion onto large black holes.  The green region indicates the parameter values allowing the formation of superheavy black hole seeds.The solid straight line marks the parameter values where $f(M_{cr})=1$, so the parameter space below this line is excluded by Eq.~(\ref{Mbound3}). }
\label{constraints}
\end{figure}

The model is fully characterized by the parameters $\xi=\sigma/H_i^3$ and $H_i/M_{pl}$, where $H_i$ is the expansion rate during inflation.  The nucleation rate of domain walls $\lambda$ depends only on $\xi$,
\beq
\lambda \sim \xi^2 e^{-2\pi^2 \xi}
\eeq
(see Eqs.~(\ref{lambdawall}), (\ref{Swall}), (\ref{A})).
The parameter space $\{\xi,H_i/M_{pl}\}$ is shown in Fig. \ref{constraints}, with red and purple regions indicating parameter values excluded by the constraints (\ref{Mbound1}) and (\ref{Mbound2}), respectively.  We show only the range $\xi\gtrsim 1$, where the semiclassical tunneling calculation is justified.  Also, for $\xi\gg 1$ the nucleation rate $\lambda$ is too small to be interesting, so we only show the values $\xi \sim$ few.

The dotted lines in the figure indicate the values $M_{cr}\sim 10^{15} ~g \equiv M_{evap}$ and $M_{cr} \sim 10^3 M_\odot \equiv M_{accr}$.  These lines, which mark the transitions between the subcritical and supercritical regimes in the excluded red and purple regions, are nearly vertical.  This is because $\xi\sim 1$ in the entire range shown in the figure, and therefore $M_{cr}\sim \xi^{-1} (M_{pl}/{H_i})^3 M_{pl}$ depends essentially only on $H_i$.  

The horizontal upper boundaries of the red and purple regions correspond to the regime of $M>M_{cr}$, where the mass function (\ref{beta2}) depends only on $\lambda$.  The corresponding constraints on $\lambda$ can be easily found: $\lambda \lesssim 10^{-15}$ for the red region (evaporation bound) and $\lambda \lesssim 10^{-26}$ for the purple region (accretion bound).  As we have emphasized, the mass function (\ref{beta2}) that we obtained for super-critical domain walls represents an upper bound on the black hole mass function, and thus the red and purple regions in Fig. \ref{constraints} may overestimate the actual size of the excluded part of the parameter space.  We have verified that the regime $t_\sigma > t_{eq}$ does not yield any additional constraints on model parameters.  

A very interesting possibility is that the primordial black holes could serve as seeds for supermassive black holes observed at the galactic centers.  The mass of such seed black holes should be $M>10^3 M_\odot$ \cite{Duechting} and their number density $n_0(M)$ at present should be comparable to the density of large galaxies, $n_G \sim 0.1 Mpc^{-3}$.  
The region of the parameter space satisfying this condition is marked in green in Fig. \ref{constraints}. 
(As before, the horizontal upper boundary of this region may overestimate the black hole mass function.)

The solid straight line in Fig. \ref{constraints} marks the parameter values where $f(M_{cr})=1$, so the parameter space below this line is excluded by Eq.~(\ref{Mbound3}).  This line lies completely within the red and purple excluded regions, suggesting that the bound (\ref{Mbound3}) does not impose any additional constraints on the model parameters.  We note, however, that the observationally excluded regions are plotted in Fig. \ref{constraints}  assuming that the expansion rate $H_i$ remains nearly constant during inflation.  Small domain walls that form black holes with masses $M_{evap}\sim 10^{15} M_\odot$ are produced towards the very end of inflation, when the expansion rate may be substantially reduced.  
The tunneling action (\ref{Swall}) depends on $H_i$ exponentially, so any decrease in $H_i$ may result in a strong suppression of small wall nucleation rate.  The evaporation bound on the model parameters can then be evaded, while larger black holes that could form dark matter or seed the supermassive black holes may still be produced.  For example, with $\xi\approx 2$, a change in $H_i$ by 20\% would suppress the nucleation rate by 10 orders of magnitude.

The black holes that form dark matter in our scenario can have masses in the range $10^{15} g < M_{bh} < 10^{24} g$.  For $M_{bh}<10^{15}g$ the black holes would have evaporated by now, and for $M_{bh} > 10^{24}g$ the line $f(M_{cr})=1$ in Fig. \ref{constraints} enters the purple region excluded by the gas accretion constraint.\footnote{The allowed mass range of black hole dark matter is likely to be wider in models where black holes are formed by vacuum bubbles.}  We note that 
Ref.~\cite{Frampton} suggested an interesting possibility that dark matter could consist of intermediate-mass black holes with $10^2 M_\odot \lesssim M_{bh} \lesssim 
10^5 M_\odot$.  Most of this range, however, appears to be excluded by the gas accretion constraint (\ref{Mbound2}).

We note also the recent paper by Pani and Loeb \cite{Pani} suggesing that
small black holes could be captured by neutron stars and could cause the
stars to implode. They argue that this mechanism excludes black holes with
masses $10^{17} g \lesssim M_{bh} \lesssim 10^{23} g$ from being the
dominant dark matter constituents.  This conclusion, however, was questioned
in Ref. \cite{Capela}.

\section{Summary and discussion}\label{conclusions}

We have explored the cosmological consequences of a population of spherical domain walls and vacuum bubbles which 
may have nucleated during the last ${\cal N}\sim 60$ e-foldings of inflation. At the time $t_i$ when inflation ends, the sizes
of these objects would be distributed in the range $t_i\lesssim R_i \lesssim e^{\cal N} t_i$, and
the number of objects within our observable universe would be of the order  $N \sim \lambda e^{3{\cal N}}$, where $\lambda$ is the dimensionless nucleation rate per unit Hubble volume.
We have shown that such walls and bubbles may result in the formation of black holes with a wide spectrum of masses.  Black holes having mass above a certain critical value would have a nontrivial spacetime structure, with a baby universe connected by a wormhole to the exterior FRW region.

The evolution of vacuum bubbles after inflation depends on two parameters, $t_b \sim (G\rho_b)^{-1/2}$ and $t_\sigma \sim (G\sigma)^{-1}$, where $\rho_b>0$ is the vacuum energy inside the bubble and $\sigma$ is the bubble wall tension. For simplicity, throughout this paper we have assumed the separation of scales $t_i \ll t_b,t_\sigma$. At the end of inflation, the energy of a bubble is equivalent to the 
``excluded" mass of matter which would fit the volume occupied by the bubble, $M_i = (4\pi/3)\rho_m(t_i)R_i^3$, where $\rho_m$ is the matter density. This energy is mostly in the form of kinetic energy of the bubble walls, which are expanding into matter with a high Lorentz factor.  Assuming that particles of matter cannot penetrate the bubble and are reflected from the bubble wall, this kinetic energy is quickly dissipated by momentum transfer to the surrounding matter on a time-scale much shorter than the Hubble time, so the wall comes to rest with respect to the Hubble flow. In this process, a highly relativistic and dense exploding shell of matter is released, while the energy of the bubble is significantly reduced. 

If the bubble is surrounded by pressureless matter, it subsequently decouples from the Hubble flow and collapses to form a black hole of mass
\begin{equation}
M_{bh} \sim \left({t_i^2\over t_b^2} + {t_i\over t_\sigma}\right) M_i \ll M_i ,
\end{equation}
which is sitting in the middle of an empty cavity.
The fate of the bubble depends on whether its mass is larger or smaller than a certain critical mass, which can be estimated as 
\begin{equation}
M_{cr} \sim {\rm Min}\{t_b,t_\sigma\}/G.
\end{equation}
For $M_{bh}<M_{cr}$, the bubble collapses into a Schwarzschild singularity, as in the collapse of usual matter. However, for $M_{bh}>M_{cr}$ the bubble avoids the singularity and starts growing exponentially fast inside of a baby universe, which is connected to the parent FRW by a wormhole. This process is represented in the causal diagram of Fig. \ref{babymultiverse}, of which a spatial slice is depicted in Fig. \ref{vb}. 
The exponential growth of the bubble size can be due to the internal vacuum energy of the bubble, or due to the repulsive gravitational field of the bubble wall. The dominant effect depends on whether $t_b$ is smaller than $t_\sigma$ or vice-versa.

If the bubble is surrounded by radiation, rather than pressureless dust, there is one more step to consider. After the bubble wall has transfered its momentum to the surounding matter and comes to rest with respect to the Hubble flow, the effects of pressure may still change the mass of the resulting black hole.
We have argued that for $M_{bh}\ll M_{cr}$ the black hole mass, given by Eq. (\ref{msth}), is not significantly affected by the pressure of ambient matter. 
However, for $M_{bh} \gg M_{cr}$ this may be important, since some radiation may follow the bubble into the baby universe, thus affecting the value of the resulting black hole mass. This process seems to require a numerical study which is beyond the scope of the present work. On general grounds, however, we pointed out that the mass of the black hole is bounded by $M_{bh} \lesssim t_H/G$, where $t_H$ is the time when the co-moving region corresponding to the initial size $R_i$ crosses the horizon.


The case of domain walls is somewhat simpler than the case of bubbles, since only the scale $t_\sigma$ is relevant. 
If the wall is small, with a horizon crossing time $t_H < t_\sigma$, and assuming that the only interaction with matter is gravitational, the wall will collapse to a black hole of mass $M_{bh}\sim \sigma t_H^2$. 
Alternatively, if the wall is impermeable to matter, a pressure supported bag of matter will form, with a substantially larger mass which is given in Eqs. (\ref{massbagdust}) or (\ref{massbagradiation}). 
These estimates apply respectively to the case where matter is non-relativistic or relativistic at the time when the wall crosses the horizon. 
Larger walls with $t_H>t_\sigma$ will begin to inflate at $t\sim t_\sigma$, developing a wormhole structure.  
If at $t\sim t_\sigma$ the universe is dominated by nonrelativistic matter, the mass of the resulting black hole is $M_{bh}\sim t_H/G$, where $t_H$ is now the time when the comoving region affected by wall nucleation comes within the horizon.  For $t_\sigma$ in the radiation era, the estimate of the mass is more difficult to obtain, since some radiation may flow into the baby universe following the domain wall. This 
issue requires a numerical study and is left for further research. As in the case of bubbles, for this case we were able to find only an upper bound on the mass, $M_{bh}\lesssim t_H/G$.

A systematic analysis of all possible scenarios would require numerical simulations, and we have not attempted to do it in this paper.  Nonetheless, for illustration, we have 
considered the case of a distribution of domain walls which interact with matter only gravitationally. Also, awaiting confirmation from a more detailed numerical study, we have assumed that large walls entering the horizon in the radiation era, with $t_\sigma \lesssim t_H \lesssim t_{eq}$, lead to black holes which saturate the upper bound on the mass, $M_{bh} \sim t_H/G$. Under these assumptions, we have shown that black holes produced by nucleated domain walls can have a significant impact on cosmology.  Hawking evaporation of small black holes can produce a $\gamma$-ray background, and radiation emitted by gas accreted onto large black holes can induce distortions in the CMB spectrum.  A substantial portion of the parameter space of the model is already excluded by present observational constraints, but there is a range of parameter values that would yield large black holes in numbers sufficient to seed the supermassive black holes observed at the centers of galaxies.  For certain parameter values the black holes can also play the role of dark matter.

Our preliminary analysis provides a strong incentive to consider the scenario of black hole formation by vacuum bubbles, which has a larger parameter space.
Note that the black hole density resulting from domain wall nucleation is appreciable only if $\xi = \sigma/H_i^3 \sim 1$, where $H_i$ is the expansion rate during inflation.  This calls for a rather special choice of model parameters, with\footnote{On the other hand, this relation may turn out to be natural in certain particle physics scenarios, once environmental selection effects are taken into account, see e.g. \cite{Yanagida}. 
We thank Tsutomo Yanagida for pointing this out to us.}
$\sigma\sim H_i^3$.  The narrowness of this range is related to the fact that the nucleation rate of domain walls is exponentially sensitive to the wall tension. 

On the other hand, the bubble nucleation rate is highly model-dependent, and the parameter space yielding an appreciable density of black holes is likely to be increased.  We note also that string theory suggests the existence of a vast  landscape of vacua \cite{Susskind}, so one can expect a large number of bubble types, some of them with relatively high nucleation rates.  

\begin{figure}
 \includegraphics[width=0.6\textwidth]{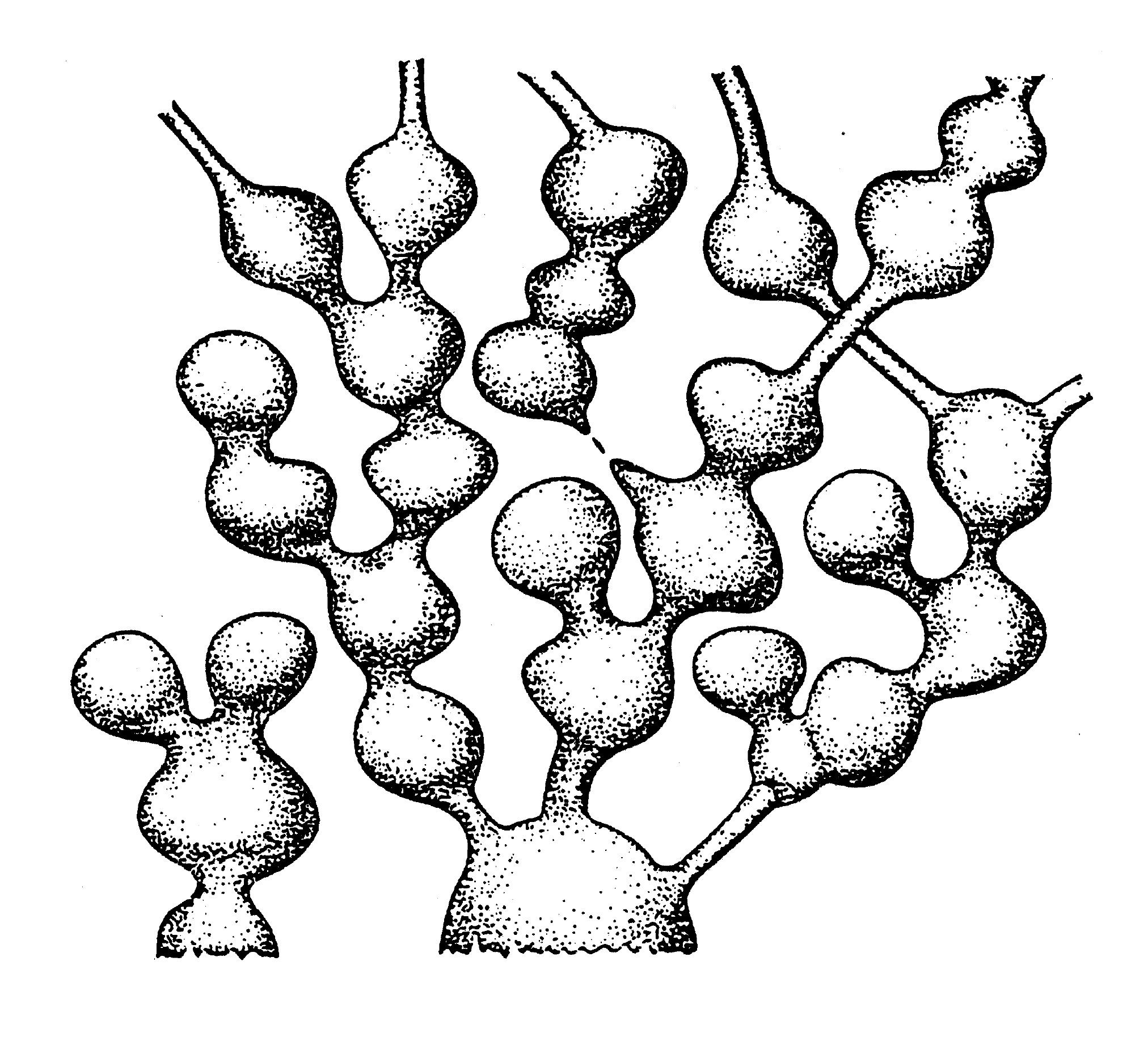}
\caption{ The eternally inflating multiverse generally has a very nontrivial spacetime structure, with a multitude of eternally inflating regions connected by wormholes.
This is reminiscent of a well known illustration of the inflationary multiverse by Andrei Linde, which we reproduce here by courtesy of the author (see also \cite{andrei} and references therein). 
In the present context the links between the balloon shaped regions should be interpreted 
as wormholes which are created by supercritical bubbles or domain walls.}\label{andrei}
\end{figure} 

Super-critical black holes produced by vacuum bubbles have inflating vacuum regions inside.  These regions will inflate eternally and will inevitably nucleate bubbles of the high-energy parent vacuum with the expansion rate $H_i$, as well as bubbles of all other vacua allowed by the underlying particle physics \cite{CdL,LeeWeinberg,recycling,andrei}.  In some of the bubbles inflation will end, and bubbles nucleated prior to the end of inflation will form black holes with inflating universes inside.  A similar structure would arise within the supercritical black holes produced by domain walls. Indeed, inflating domain walls can excite any fields coupled to them, even if the coupling is only gravitational. In particular, they can excite the inflaton field, causing it to jump back to the inflationary plateau \cite{GBGM,M}, or they may cause transitions to any of the accessible neighboring vacua \cite{Czech}. We thus conclude that the eternally inflating multiverse generally has a very nontrivial spacetime structure, with a multitude of eternally inflating regions connected by wormholes.\footnote{A similar spacetime structure was discussed in a different context by Kodama et al \cite{Kodama}.}
This is reminiscent of a well known illustration of the inflationary multiverse, by Andrei Linde
which we reproduce in Fig. \ref{andrei}. 
In the present context the links between the balloon shaped regions should be interpreted as wormholes. We note
that the mass distributions of black holes resulting from domain walls and from vacuum bubbles are expected to be different and can in principle be distinguished observationally.

Black hole formation in the very early universe has been extensively studied
in the literature, and a number of possible scenarios have been proposed
(for a review see \cite{Carr2,Khlopov}).  Black holes could be formed at
cosmological phase transitions or at the end of inflation.  All these
scenarios typically give a sharply peaked mass spectrum of black holes, with
the characteristic mass comparable to the horizon mass at the relevant
epoch.  In contrast, our scenario predicts a black hole distribution with a
very wide spectrum of masses, scanning many orders of magnitude.\footnote{It
has been recently suggested \cite{Juan} that black holes with a relatively
wide mass spectrum could be formed after hybrid inflation \cite{JuanLinde}.
However, the spectrum does not typically span more than a few orders of
magnitude, and its shape is rather different from that in our model.}  If a
black hole population with the predicted mass spectrum is discovered, it
could be regarded as evidence for inflation and for the existence of a
multiverse. 

We would like to conclude with two comments concerning the global structure of supercritical black holes, which may be relevant to the so-called black hole information paradox (see Ref. \cite{daniel} for a recent review). First of all, given that the black hole has a finite mass, the amount of information carried by Hawking radiation as the black hole evaporates into the asymptotic FRW universe will be finite. For that reason, it cannot possibly encode the quantum state of the infinite multiverse which develops beyond the black hole horizon. Second, the supercritical collapse leads to topology change, and to the formation of a baby universe which survives arbitrarily far into the future. In this context, it appears that unitary evolution should not be expected in the parent universe, once we trace over the degrees of freedom in the baby universes. Finally, we should stress that the formation of  wormholes, leading to topology change and to formation of baby multiverses within black holes, does not require any exotic physics. It occurs by completely natural causes from regular initial conditions.\footnote{This does not contradict the result of Farhi and Guth \cite{FG}, that a baby universe cannot be created by classical evolution in the laboratory. That result follows from the existence of an anti-trapped surface in the inflating region of the baby universe, the null energy condition, and the existence of a non-compact Cauchy surface. With these assumptions, Penrose's theorem \cite{Roger} implies that some geodesics are past incomplete. This is a problem if our laboratory is embedded in an asymptotically Minkowski space, which is geodesically complete. However, it is not a problem in a cosmological setting which results from an inflationary phase. Inflation itself is geodesically incomplete to the past \cite{BGV}, if we require the inflationary Hubble rate to be bounded below. In the present context, such geodesic incompleteness is not related to a pathology of baby universes, but it is just a feature of the inflationary phase that precedes them. It should also be noted that the infinite Cauchy surface in the FRW universe of interest
is not necessarily a global Cauchy surface.  For example, the spacetime to the past of the end-of-inflation surface $t=t_i$ could be described by a closed chart of de Sitter
space, in which case all global Cauchy surfaces would be compact and Penrose's theorem would not be applicable.}

\section{Acknowledgements}

This work is partially supported by MEC FPA2013-46570-C2-2-P, AGAUR 2014-SGR-1474, CPAN
CSD2007-00042 Consolider-Ingenio 2010 (JG) and by the National Science Foundation (AV and JZ).  J.Z. was also supported by the Burlingame Fellowship at Tufts University.
A.V. is grateful to Tsutomu Yanagida for a useful discussion, suggesting the possible relevance of our scenario for dark matter and to Abi Loeb for several useful discussions.

\appendix

\section{Evolution of test domain walls in FRW}

In this appendix, we solve the evolution of test domain walls in a FRW universe numerically and find the numerical coefficient $C_r$ and $C_{m}$ defined in Sec. III. 

The metric for a FRW universe is
\ba
ds^2= a^2(\eta)(-d\eta^2 + dr^2+r^2 d\Omega^2),
\ea
where $\eta$ is the conformal time. The scale factor is given by
\ba
a(\eta) = \left(\frac{\eta}{\eta_i}\right)^{p},
\ea
where $p=2$ for the dust dominated background and $p=1$ for the radiation dominated background. We choose $\eta_i = p$, so that $H_i = 1$.

The domain wall worldsheet is parametrized by $\eta$, and the worldsheet metric can be written as
\ba
ds_3^2 =a^2(\eta) \left[\left(-1+ r'^2\right)d\eta^2 + r^2d\Omega^2\right].
\ea
In this Appendix, prime indicates a derivative with respect to the conformal time $\eta$. The domain wall action is proportional to the worldsheet area,
\ba
S= 4\pi \sigma \int d\tau \sqrt{1- r'^2} a^3 r^2,
\ea
and the equation of motion is
\ba
r'' + 3 \frac{a'}{a} r' (1-r'^2)+\frac{2}{r}(1-r'^2)=0.
\ea
We solve this equation with the initial conditions
\ba
r(\eta_i)=R_i, ~~~ r'(\eta_i)=0.
\ea
The solutions are plotted in Fig. \ref{rt} in terms of the cosmological time $t$.

We regard a black hole as formed when the radial coordinate drops to $r < 10^{-4} t_i$ and determine its mass from Eq.~(\ref{mwall}), which in terms of the conformal time $\eta$ takes the form
\ba
M (\eta) = 4\pi\sigma \frac{ a^2 r^2}{\sqrt{1-r'^2}}.
\ea
This mass grows while the wall is being stretched by Hubble expansion, but remains nearly constant during the late stages of the collapse.

\begin{figure}
\includegraphics[width=0.45\textwidth]{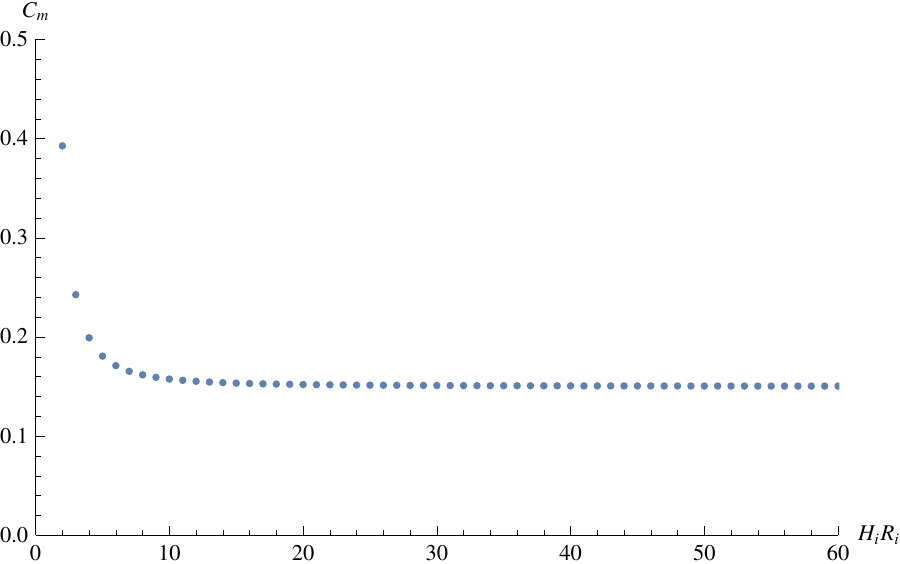}
\includegraphics[width=0.45\textwidth]{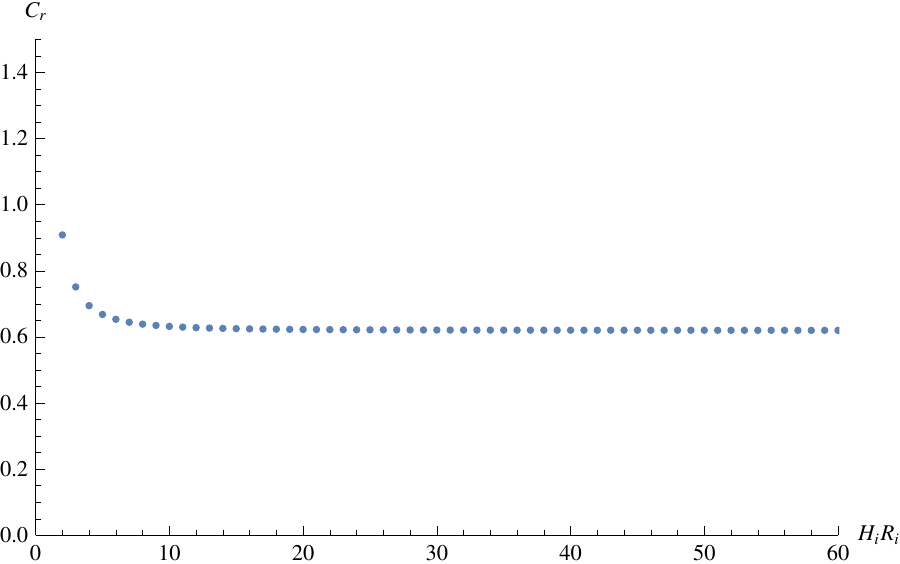}
\caption{The dependence of $C_{m}$ and $C_r$ on $H_i R_i=R_i$. For $H_iR_i > 10$, we find $C_{m}\simeq 0.15$ and $C_r \simeq 0.62$.}
\label{fig:c}
\end{figure} 

According to Sec. III, the numerical coefficients  $C_m$ and $C_r$ are defined as
\ba
C_{m,~r} = \frac{p^2}{(1+p)^2}\frac{M_{bh}}{4\pi \sigma  {t}_{H}^2},
\ea
where the Hubble-crossing time ${t}_{H}$ satisfies 
\ba
R_i a({t}_{H}) = \frac{1+p}{p}{t}_{H},
\ea
The dependence of $C_{m,r}$ on $H_iR_i = R_i$ is shown in Fig.~\ref{fig:c}.  We see that the coefficients approach the values $C_{m}\simeq 0.15$ and $C_r \simeq 0.62$ for $H_i R_i > 10$.

\section{Large domain wall in dust cosmology}\label{appendixglobal}

In this Appendix we describe in more detail the construction of the solution represented in Fig. \ref{largewall}, which corresponds to a large domain wall embedded in a dust cosmology.
The gravitational field of the wall creates a wormhole structure leading to a baby universe. We shall start by cosidering the matching of Schwarzschild to a dust cosmology, and then the 
matching of two Schwarzschild metrics accross a domain wall. Finally, the different pieces are put together.

\subsection{Matching Schwarzschild to a dust cosmology}

\begin{figure}
 \includegraphics[width=1\textwidth]{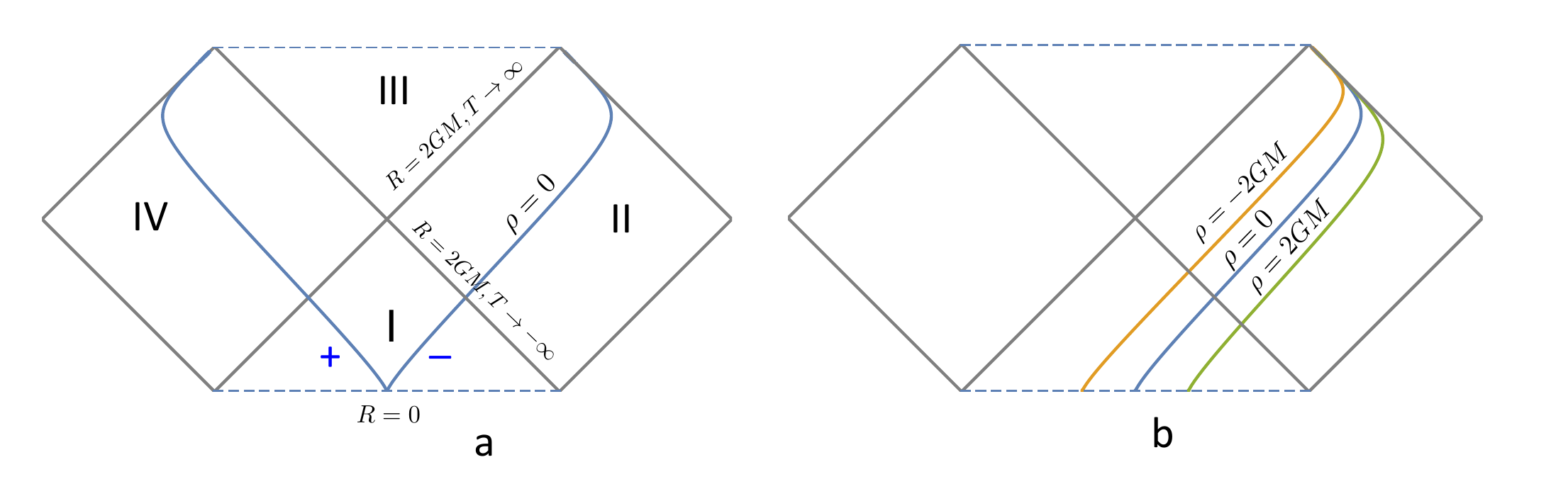}
\caption{Lemaitre coordinates in the extended Schwarzschild diagram. The left panel shows the curves $\rho=0$ for the two possible choices of the double sign in Eq.(\ref{double}).
In the left panel, we adopt the lower sign, indicated by $-$, which corresponds to radial geodesics traveling towards region II from the white hole singularty at $R=0$, for 
different values of $\rho$. Note that such geodesic coordinates only cover half or the extended Schwarzschild solution.
 }
\label{Kruskal}
\end{figure}

Consider the Schwarzschild metric 
\begin{equation}
ds^2=-\left(1-{2GM\over R}\right) dT^2 +  \left(1-{2GM\over R}\right)^{-1} dR^2 + R^2 d\Omega^2.
\end{equation}
In Lemaitre coordinates, this takes the form
\begin{equation}
ds^2 = -d\hat t^2 + {2GM\over R} d\rho^2 + R^2 d\Omega^2. \label{syn}
\end{equation}
where $\hat t$ and $\rho$ are defined by the relations
\begin{eqnarray}
d\hat t &=& \pm dT - \sqrt{2GM\over R}  \left(1-{2GM\over R}\right)^{-1} dR \label{dhatt}, \\ 
d\rho &=& \mp dT + \sqrt{R\over 2GM} \left(1-{2GM\over R}\right)^{-1} dR \label{drho}.
\end{eqnarray}
Adding (\ref{dhatt}) and (\ref{drho}), we have
\begin{equation}
R=\left[{3\over 2}(\hat t+\rho)\right]^{2/3}(2GM)^{1/3},\label{r} 
\end{equation}
where an integration constant has been absorbed by a shift in the origin of the $\rho$ coordinate.
The expression for $T$ as a function of $\hat t$ and $\rho$ can be found from (\ref{drho}) as:
\begin{equation}
\pm T=4GM \left({1\over 3}\left(R\over 2 GM\right)^{3/2}+ \sqrt{R\over 2GM}-\tanh^{-1}\sqrt{R\over 2GM} \right)-\rho_\pm,\label{double}
\end{equation}
with $R$ given in (\ref{r}). A second integration constant has been absorbed by a shift in the $T$ coordinate. Since the metric (\ref{syn}) is synchronous, the lines of constant spatial coordinate are geodesics, and 
$\hat t$ is the proper time along them. The double sign in front of $T$ corresponds to two possible choices of geodesic coordinates $\rho_\pm$.
We shall consider geodesics which start from the singularity at $R=0$ at the bottom of the Kruskal diagram, see Fig. \ref{Kruskal}. In the region $R<2GM$ 
the $T$ coordinate may grow or decrease depending on which spatial direction we go, while $\hat t$ and $R$ both grow towards the future. 
A sphere of test particles at $\rho=\rho_0$ has the physical radius given by 
\begin{equation}
R
\propto (\hat t+\rho_0)^{2/3}.
\end{equation}
Thus, co-moving spheres with different values of the $\rho$ coordinate have a similar behaviour, with a shifted value of proper time.

The metric (\ref{syn}) can be matched to a flat FRW metric
\begin{equation}
ds^2 = -d t^2 + a^2(t)  (d r^2 +  r^2 d\Omega^2).
\end{equation}
The co-moving sphere with $\rho=\rho_0$ will be glued to a co-moving sphere with $r= r_0$. Continuity of the temporal and
angular components of the metric accross that surface requires 
\begin{equation}
R =  a(t)\,r_0,
\end{equation}
with $t = \hat t+\rho_0$, and
\begin{equation}
 a(t)\propto t^{2/3},
\end{equation}
corresponding to a dust cosmology.  Such cosmology has a matter density given by $\rho_m=1/(6\pi G t^2)$. Using (\ref{r}) with $\rho=\rho_0$, it is then straightforward to show that
\begin{equation}
M={4\pi\over 3} \rho_m\,R^3. \label{relmass}
\end{equation}
Therefore, the mass parameter $M$ in the Schwarzschild solution corresponds to the total mass of dust that would be contained within the sphere of radius $R$,  where 
the two solutions are matched. The relation (\ref{relmass}) can also be derived by imposing that the dust particle at the edge of the FRW metric should be a geodesic of both
FRW and Schwarzschild. This is explained in Subsection \ref{largewallindust}.

Since the normal $n^\mu$ to the the matching hypersurface $\rho=const.$ or $r=const.$ has only radial component, and the metric is diagonal, the extrinsic curvature 
$K_{\mu\nu}=n_{\mu;\nu}$ takes a simple form given by
\begin{equation}
K_{\alpha\beta} = {1\over 2}\partial_n g_{\alpha\beta}
\end{equation}
where $\partial_n = a^{-1}\partial_r = (R/2GM)^{1/2} \partial_\rho$ is the normal derivative, and $\alpha$ and $\beta$ run over the temporal and angular coordinates.
It is straightforward to check that $K_{\alpha\beta}$ has only angular components, and that these are the same on both sides of the matching hypersurface:
\begin{equation}
K_{\Omega\Omega'} = g_{\Omega\Omega'}/R.
\end{equation}
Here $\Omega$ and $\Omega'$ run over the angular coordinates. The continuity of the metric and of the extrinsic curvature means that there is no distributional source at the junction.

\begin{figure}
 \includegraphics[width=1\textwidth]{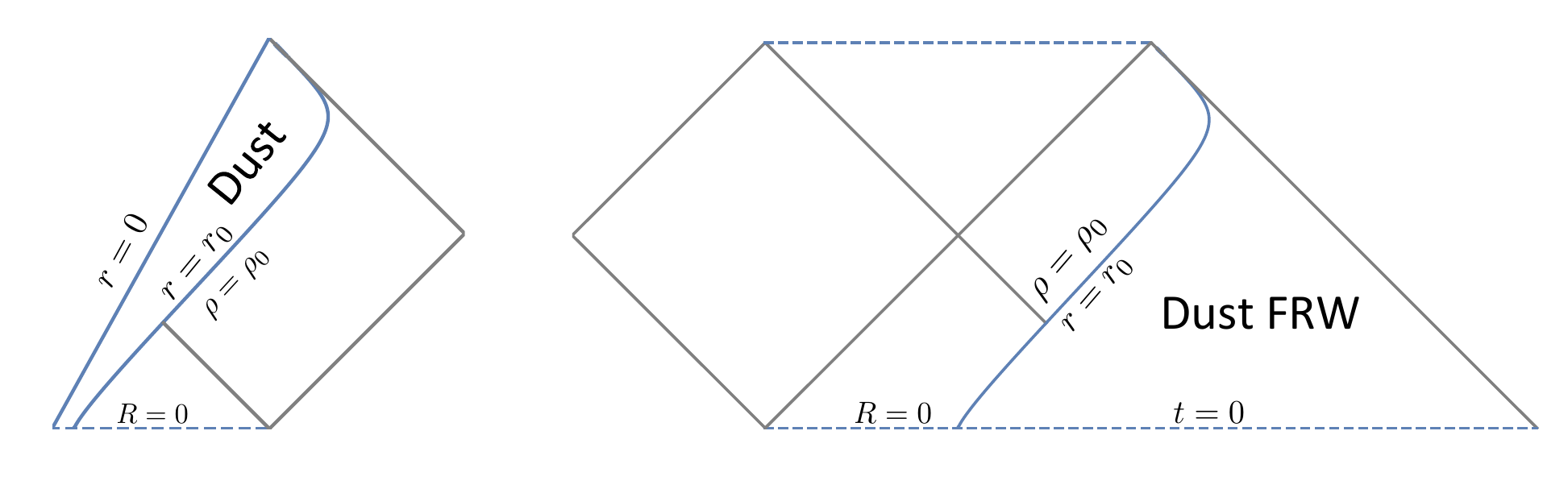}
\caption{A Schwarzschild solution can be matched to a dust FRW solution at ta $\rho=\rho_0$ hypersurface. This  can be done in two different ways. The left panel shows an expanding ball of dust matched to a Schwarzschild interior, while the right panel shows a black hole interior embedded in a dust FRW exterior. 
 }
\label{matching}
\end{figure}

The matching of Schwarzshild and FRW metrics can be done in two different ways, which are represented in Fig. \ref{matching}. An expanding ball of matter may be embedded in a Schwarzschild exterior solution,
as indicated in the left panel, or a black hole interior can be embedded in a dust FRW exterior, as indicated in the right panel. In this case, the black hole connects through a wormhole to the asymptotically flat region $IV$.

\subsection{Domain wall in Schwarzschild}

\begin{figure}
 \includegraphics[width=1\textwidth]{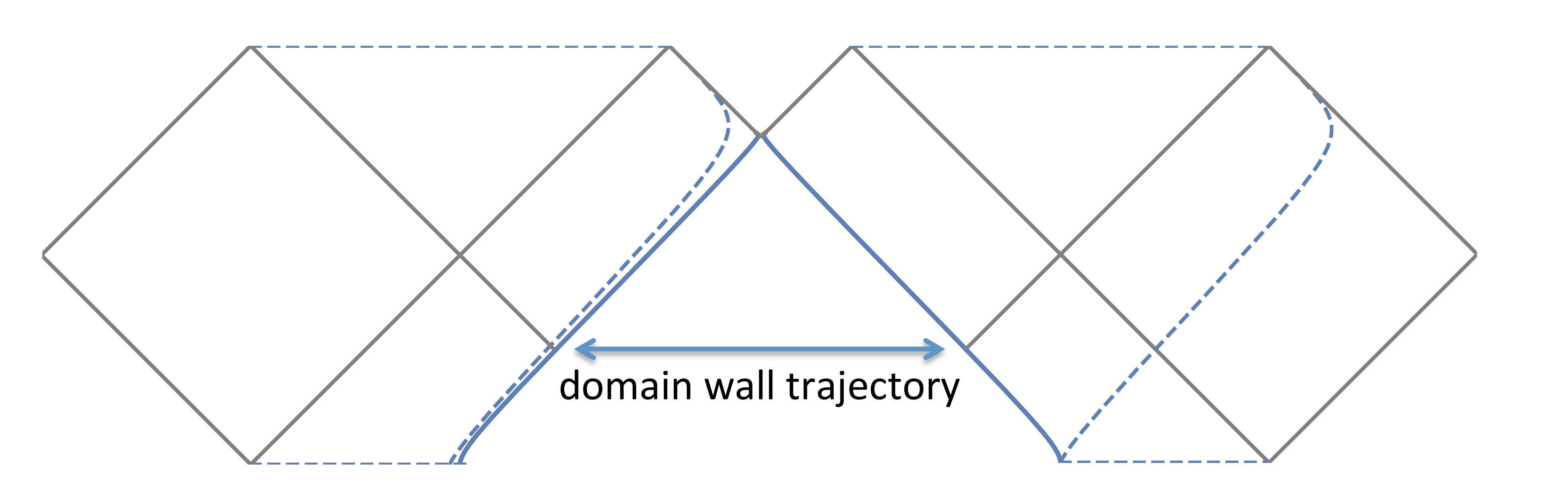}
\caption{The trajectory of a domain wall between two segments of Schwarzschild solutions with the same mass $M>M_{cr}=t_\sigma/(3\sqrt{3}G)$.
 }
\label{largewallSchwarzschild}
\end{figure}

Let us now consider the motion of a domain wall in a spherically symmetric vacuum. In Lemaitre coordinates, a radial vector which is tangent to the worldsheet is given by 
$u^\mu=(\dot{\hat t}, \dot \rho,0,0)$, where a dot indicates derivative with respect to the wall's proper time $\tau$: 
\begin{equation}
\dot\rho = {d\rho\over d\tau}.
\end{equation}
The normal to the worldsheet is then given by
\begin{equation}
n_\mu = \sqrt{2GM\over R}\left(-\dot\rho, \dot{\hat t},0,0\right).
\end{equation}
Since the normal is in the radial and temporal directions, the angular components of the extrinsic curvature $K_{\mu\nu} = n_{\mu;\nu}$ are given by the simple expression
\begin{equation}
K_{\Omega\Omega'} = {1\over 2} n^\mu\partial_\mu g_{\Omega\Omega'}=\left({2GM\over R}\dot\rho +\dot{\hat t}\right){g_{\Omega\Omega'}\over R}.\label{extrinsic}
\end{equation}
The normalization of the radial tangent vector $u^\mu$ requires that
\begin{equation}
\dot{\hat t} = \sqrt{1+{2GM\over R}\dot\rho^2}.\label{dotau}
\end{equation}
Israel's matching conditions for a thin domain wall of tension $\sigma$ imply that the extrinsic curvature should change by the amount
\begin{equation}
[K_{\Omega\Omega'}] = -{2\over t_\sigma} g_{\Omega\Omega'},\label{israel}
\end{equation}
as we go accross the wall in the positive $\rho$ direction.
Here 
\begin{equation}
t_\sigma=(2\pi G\sigma)^{-1}.
\end{equation}
By continuity of the metric accross the wall, the proper radius of the wall as a function of proper time, $R(\tau)$, should be the same as calculated from both sides. 
This requires that the trajectory of the wall as seen from the ``inside", $\rho_-(\hat t_-)$, should be glued to a mirror image of this trajectory, $\rho_+(\hat t_+)$. Here, the double sign $\pm$ refers to the choice of Lemaitre coordinate system in Eq. (\ref{double}), see Fig. \ref{Kruskal}, but $\rho_+(\hat t_+)$ and $\rho_-(\hat t_-)$ have the same functional form, which we shall simply denote by $\rho(\hat t)$. The extrinsic curvature of the domain wall will be the same from both sides, but with opposite sign.
Combining (\ref{extrinsic}), (\ref{dotau}) and  (\ref{israel}), we have
\begin{equation}
{2GM\over R}\dot\rho+ \sqrt{1+{2GM\over R}\dot\rho^2}={R\over t_\sigma},
\end{equation}
which has the solution
\begin{equation}
\dot\rho= \left(1-{2GM\over R}\right)^{-1}\left[-{R\over t_\sigma} + \sqrt{{R\over 2GM}\left({R^2 \over t_\sigma^2}+{2GM\over R}-1\right)}\right].
\end{equation}
The trajectory of a domain wall gluing two segments of the Schwarzschild solution is depicted in Fig. \ref{largewallSchwarzschild}.

Note that for $R=t_\sigma$ we have $\dot\rho=0$, and it is straightforward to check that
\begin{eqnarray}
\dot\rho>0,  &{\rm for}& R> t_\sigma, \label{bigger}\\
\dot\rho< 0, &{\rm for}& R< t_\sigma,
\end{eqnarray}
a property which will be important for the following discussion.

A similar calculation in the Schwarzschild coordinates gives a simpler form for the equations of motion \cite{Blau},
\begin{eqnarray}
\dot R^2 &=&-V(R), \label{uni}\\
\dot T &=& \pm\left(1-{2GM\over R}\right)^{-1}{R\over t_\sigma}.
\end{eqnarray}
where
\begin{equation}
V(R)=1-{2GM\over R}-\left({R\over t_\sigma}\right)^2.
\end{equation}
Eq. (\ref{uni}) is analogous to the equation of motion for a non-relativistic particle of zero energy in a static potential $V(R)$.
The maximum of this potential is at 
\begin{equation}
R_m=\left({GM t_\sigma^2}\right)^{1/3}.
\end{equation}
The value of $V(R_m)$ is negative (or positive) for $M$ larger (or smaller) than a critical mass $M_{cr}$ given by:
\begin{equation}
M_{cr} = {t_\sigma \over 3\sqrt{3} G}.
\end{equation}
For $M<M_{cr}$, a domain wall which starts its expansion at the white hole singularity will bounce at some $R<R_{m}$ and recollapse at a
black hole singularity. Here we are primarily interested in the opposite case $M>M_{cr}$, where the expansion of the wall radius is unbounded and
continues past the maximum of the potential at $R_m$, towards infinite size (see Fig. \ref{largewallSchwarzschild}).

 \label{gluingschw}

\subsection{Domain wall in dust}

As mentioned around Eq. (\ref{bigger}), for $R>t_\sigma$ the wall recedes from matter geodesics with the escape velocity, on both sides of the wall. Because of that, we can construct an exact solution where an expanding ball of matter, described by a segment of a flat matter dominated FRW solution is glued to a vacuum layer represented by a segment of the Schwarzschild solution. This in turn is glued to a 
second vacuum layer also described by a Schwarzschild solution of the same mass on the other side of the wall, as described in Subsection \ref{gluingschw}. Finally, the second layer is glued to an exterior flat matter dominated FRW solution. The result is represented in Fig. \ref{largewall}.

\end{document}